\documentclass[journal]{IEEEtran}
\ifCLASSINFOpdf
\else
\fi
\usepackage[pdftex]{graphicx}
\usepackage{ascmac}
\usepackage{subfigure}
\usepackage{bm}
\usepackage{amsmath}
\usepackage{amsfonts}
\usepackage{ascmac}
\usepackage{algorithm}
\usepackage{algorithmicx}
\usepackage{algpseudocode}
\usepackage{multirow}
\usepackage{url}
\usepackage{listings}

\newcommand{\prmt}[1]{\bm{\mathrm{#1}}}
\algnewcommand\algorithmicforeach{\textbf{for each}}
\algdef{S}[FOR]{ForEach}[1]{\algorithmicforeach\ #1\ \algorithmicdo}

\algdef{SE}[DOWHILE]{Do}{doWhile}{\algorithmicdo}[1]{\algorithmicwhile\ #1}%

\def\vgd{V_{\mathrm{gd}}}

\def\cds{C_{\mathrm{ds}}}
\def\cgs{C_{\mathrm{gs}}}
\def\cgd{C_{\mathrm{gd}}}

\begin{document}
%
\title{Accelerating Parameter Extraction of Power MOSFET Models
  Using Automatic Differentiation}
%
%
%

\author{Michihiro~Shintani,
        Aoi~Ueda,
        and~Takashi~Sato
        \thanks{M. Shintani and A. Ueda are with Graduate School of Science
          and Technology, Nara Institute of Science and Technology (NAIST), Japan.
        }
        \thanks{T. Sato is with Graduate School of Informatics, Kyoto
          University, Japan.}
}

%
%

\markboth{}%
{Shell \MakeLowercase{\textit{et al.}}: Bare Demo of IEEEtran.cls for IEEE Journals}
%



\maketitle


\begin{abstract}%
  The extraction of the model parameters is as important as the
  development of compact model itself because simulation accuracy is
  fully determined by the accuracy of the parameters used.  This study
  proposes an efficient model-parameter extraction method for compact
  models of power MOSFETs.  The proposed method employs automatic
  differentiation (AD), which is extensively used for training
  artificial neural networks. In the proposed AD-based parameter
  extraction, gradient of all the model parameters is analytically
  calculated by forming a graph that facilitates the backward
  propagation of errors.  Based on the calculated gradient,
  computationally intensive numerical differentiation is eliminated
  and the model parameters are efficiently optimized. Experiments are
  conducted to fit current and capacitance characteristics of
  commercially available silicon carbide MOSFET using power MOSFET
  models having 13 model parameters. Results demonstrated that the
  proposed method could successfully derive the model parameters
  3.50$\times$ faster than a conventional numerical-differentiation
  method while achieving the equal accuracy.
\end{abstract}

\begin{IEEEkeywords}
Parameter extraction, SPICE compact model, Automatic differentiation, Power MOSFET compact model
\end{IEEEkeywords}

%
\IEEEpeerreviewmaketitle

\section{Introduction}

Wide-bandgap devices, such as silicon carbide
(SiC) and gallium nitride (GaN) metal oxide semiconductor field-effect
transistors (MOSFETs), are the promising alternatives that
replace conventional silicon (Si) power
devices~\cite{Baliga_book,Kimoto_book}. Owing to the excellent
material properties, SiC and GaN MOSFETs can operate at higher
switching frequencies
with lower switching loss
under wide range of ambient temperature.  Especially, the high
switching operation greatly contributes to reducing the volume and the
weight of power converters. Those are the desirable properties in
the application of electric vehicles (EVs), such as on-board charger
(OBC) and in-wheel motor (IWM)~\cite{TIE2917_Liu,ECCE2020_Akatsu}.
As the operating frequency of the power converters increases, design
optimization using circuit simulators based on compact model of
power MOSFETs plays an increasingly important role.





The formulation of accurate compact models for power MOSETs, and the
careful extraction of their parameter sets are crucial for obtaining
reliable results from circuit simulations. Compact power MOSFET
models, which have long been the topic of intensive researches, are
composed of multiple nonlinear functions that accurately capture the
device physics of the MOSFET.  Recently, surface-potential-based and
charge-based compact
models~\cite{TED2013_Mattausch,TPEL2018_Shintani,TED2019_Agarwal,TED2020_Albrecht}
have been successfully applied to simulate power converters.

One of the most widely used parameter-extraction methods is iterative
parameter refinement, which is based on gradient
calculations~\cite{JJAP2004_Li,ICMTS2009_Zhou}.  This approach
involves two processes, gradient calculation and parameter updating,
which are repeated until the characteristics of the model agree with
the measured values, or the iteration limit is reached. In this
context, the gradient is a set of partial derivatives for all model
parameters in the model equation. Then, in the parameter-update phase,
the gradient-descent algorithm is applied to refine the parameters
based on the gradient that indicates the direction in which the
parameters should be adjusted to improve the fitting accuracy.

Although iterative parameter refinement is a very general method and is
applicable to arbitrary model equations, it tends to consume a long time for
all the parameters to converge using this approach. The gradient
calculations require numerical differentiation (ND) with regard to each
parameter, which demands full evaluation of the model equations for all bias
voltages and all model parameters. Therefore, the model evaluation
constitutes most of the time required for parameter optimization. In a
preliminary experiment~\cite{ICMTS2018_Shintani}, even when a very simple
device model equation~\cite{TED1991_sakurai} was used, 99.6\% of the
parameter optimization time was spent on the gradient calculation.

Several techniques are known to derive derivatives: examples include
symbolic differentiation (SD), numerical differentiation (ND), and
automatic differentiation (AD). SD derives derivatives based on the
symbolic rules of basic calculus.  While it provides us with exact
computation, SD tends to require a large amount of memory resource
when the model equation becomes complex. Since ND needs to evaluate
the model equation twice, the calculation time tends to be long
depending on the model scale and the number of model parameters.  AD
is a technique for deriving partial derivatives for the equations
defined in a program. AD can be carried out by the combinations of
four basic arithmetic operations and elementary functions, such as
exponential function, etc., through repeatedly applying the chain
rule.  By using AD, partial derivatives can be automatically obtained
with a reduced calculation cost.

Herein, we propose a method to accelerate the model parameter extraction for
power MOSFET models. Expanding upon the idea proposed
in~\cite{ICMTS2018_Shintani}, we implement automatic differentiation
(AD)~\cite{JCAM2000_Bartholomew-Biggs}, which has recently been extensively
applied in machine learning to adjust the weights of convolutional neural
networks during the process of {\it
  backpropagation}~\cite{JMLR2018_Baydin,NATURE2015_LeCun}. The AD technique
enables the contribution of each model parameter to the discrepancy between
the measurement and the model to be calculated individually. This is
achieved by traversing a so-called computational graph merely twice, i.e.,
once in the forward direction and once in the backward direction, regardless
of the number of model parameters. Therefore, the proposed method eliminates
the need for repeated model evaluations during the gradient calculation. The
AD can be directly substituted for the conventional ND so there is no need
to change the existing procedures for updating the parameters in the
gradient-based parameter-extraction method. Herein, we experimentally
demonstrate that this approach reduces the time required for parameter
extraction by 4.34$\times$ than the ND method for a model consisting of eight
parameters; this is close to the theoretical upper limit of 4.5$\times$ time
reduction with this number of parameters.

The remainder of this paper is organized as follows:
Section~\ref{sec:conv} provides the problem formulation of the ND. In
Section~\ref{sec:propose}, the basic concept of AD and the AD-based
parameter-extraction method are described. Then, experiments using two
SPICE models~\cite{TED1991_sakurai,TPEL2018_Shintani} to quantitatively
evaluate the effectiveness of the proposed method are reported in
Section~\ref{sec:exp}. Finally, the paper is concluded in
Section~\ref{sec:conclusion}.

\section{Parameter extraction based on numerical differentiation}
\label{sec:conv}

\subsection{Gradient-Based Parameter Extraction}
We first review gradient-based parameter extraction as the typical
optimization algorithm, while the proposed method is applicable to
other algorithms using derivatives, such as the Levenberg-Marquardt
(LM) method~\cite{SIAM1963_Marquardt}.  Algorithm~\ref{alg:gd}
outlines the parameter extraction for a drain current model,
$f(\cdot)$, as an example. Here, $f(\cdot)$ is a function of the
gate-source voltage, $V_{\rm gs}$, and drain-source voltage, $V_{\rm
  ds}$, and is based on the current model equation with constant model
parameters, $\bm p$.

The algorithm has the following seven inputs: the initial value vector of
the model parameters ($\bm p$), the vector representing the small changes in
each parameter ($\bm{\delta}$), the measured current data
($\bm{I}^{\rm meas}$) and the corresponding bias voltages ($\bm{V}$), the
target error ($E_{\rm target}$), and the maximum number of iterations
($N_{\rm max}$). $\bm{\eta}$ represents the rate at which the parameters are
updated. Vectors $\bm{p}$, $\bm{\delta}$, and $\bm{\eta}$ all have sizes of
$n$, which is equal to the number of model parameters. $\bm{V}$ is a vector
whose components represent the voltage pair $V_{\rm gs}$ and ${V_{\rm ds}}$
at which the drain current, $\bm{I}^{\rm meas}$, is measured. Vectors
$\bm{V}$ and $\bm{I}^{\rm meas}$ both have dimensions of
$m_{\rm vd} \times m_{\rm vg}$, where $m = m_{\rm vd} \times m_{\rm vg}$ is
the total number of the data points measured and $m_{\rm vd}$ and
$m_{\rm vg}$ are the number of voltages measured in the ${V_{\rm ds}}$ and
${V_{\rm gs}}$ sweeps, respectively.

The gradient-based parameter extraction proceeds by changing each of the
parameters, $\bm{p}$, in the direction determined by the numerical
derivatives to reduce the cost function, $E$. Here, $E$ is the
root-mean-square error (RMSE) between the simulation and measurement and is
generally defined as follows:
\begin{equation}
  E = \sqrt{\frac{1}{m}\sum_{j=0}^{m-1} (I^{\rm meas}_j-I^{\rm sim}_j)^2}.
  \label{eq:error}
\end{equation}
Before the extraction is initiated, the currents may be normalized such
that each bias point has an equal contribution to the final
error~\cite{TED2008_Takeuchi}.  Otherwise, the fitting results would be
dominated by the bias points having larger current values, while those
having smaller current values do not contribute to the fitting.  This
leads to insufficient extraction accuracy.
In line 3 of Algorithm~\ref{alg:gd}, the function {\em gradient\_calc}
is used to obtain the gradient of the cost function, $\nabla E$, in
terms of each parameter as follows:
\begin{equation}
  \nabla E = \left(\frac{\partial E}{\partial p_0}, \ldots ,\frac{\partial
      E}{\partial p_{n-1}}\right).
\end{equation}

Then, in lines 4-6, the gradient is used to update the parameters
according to $\bm{\eta}$ in a direction that will reduce the cost
function. In line 5, each parameter is changed according to $\partial
E/\partial p_i$ at an update rate of $\bm{\eta}$, which is assumed to be
constant throughout the parameter extraction for the sake of
simplicity. However, the variable step size can be chosen as proposed in
AdaGrad~\cite{JMLR2011_Duchi} or AdaDelta~\cite{arxiv2012_adadelta}.

Next, the cost function and its gradient are recalculated based on the
updated model parameters. The gradient calculation and the parameter
update are alternatively repeated until either of the exit conditions in
line 2 is satisfied, i.e., the cost function becomes smaller than the
target value or the iteration limit is reached.


\begin{figure}[t!]
  \begin{algorithm}[H]
    \caption{Gradient-descent-based parameter optimization}
    \label{alg:gd}
    \begin{algorithmic}[1]
      \Require $\bm{p}=(p_0, \ldots ,p_{n-1})$,
      $\bm{\delta}=(\delta_0, \ldots ,\delta_{n-1})$,
      ${\bm V}=((V_{\rm{gs}_0},
      V_{\rm {ds}_0}), \ldots ,(V_{\rm{gs}_{m-1}},V_{\rm{ds}_{m-1}}))$,
      $\bm{I}^{\rm meas}=(I^{\rm meas}_0, \ldots ,I^{\rm meas}_{m-1})$,
      $E_{\rm target}$, $N_{\rm max}$,
      $\bm{\eta}=(\eta_0, \ldots, \eta_{n-1})$
      \State initialize $N_{\rm iter}=0$
      \Do
      \State $\nabla E, E =$ gradient\_calc($\bm{p}$, $\bm{\delta}$,
      ${\bm V}$, $\bm{I}^{\rm meas}$)
      \ForEach {$p_i \in \bm{p}$}
      \State $p_i = p_i - \eta_i \frac{\partial E}{\partial p_i}$
      \EndFor
      \State $N_{\rm iter}$++
      \doWhile{($N_{\rm iter} < N_{\rm max}$ or $E_{\rm target} < E$) }
      \State Return optimal parameter $\bm{p}$
    \end{algorithmic}
  \end{algorithm}
\end{figure}

\subsection{Numerical Differentiation}\label{sec:nd}

The gradient-based parameter extraction requires the gradient to be
calculated as shown in line 3 of Algorithm~\ref{alg:gd}.  A simple way
to approximate the derivatives is to adopt ND; in this approach, one of
the model parameters, $p_i$, is slightly changed by a small amount,
$\delta_i$, while the other model parameters and inputs are fixed to
evaluate the change in $E$. This two-point gradient approximation is
versatile as it can be applied even when the model equation is not
expressed using closed-form equations.

The ND-based gradient calculation is conducted by the function {\em
gradient\_calc} in Algorithm~\ref{alg:gd} as summarized in
Algorithm~\ref{alg:nd}. The ND approximates the derivative based on the
slope between the two points; thus, the model equation, $f(\cdot)$, must
be evaluated twice (i.e., once with respect to each model parameter,
$I_{\rm sim1}$ and $I_{\rm sim2}$). $E$ and $E_{\rm delta}$ are the
RMSEs for $\bm{p}$ and $\bm{p}'$, respectively. Based on these errors,
the partial derivative of $E$ is calculated with respect to each
parameter as shown in line 14 of Algorithm~\ref{alg:nd}.

\begin{figure}[t!]
  \begin{algorithm}[H]
    \caption{ND-based gradient calculation}
    \label{alg:nd}
    \begin{algorithmic}[1]
      \Function{ND}{$\bm{p}$, $\bm{\delta}$, ${\bm V}$, $\bm{I}^{\rm meas}$}
        \State $e=0$, $e_{\rm delta}=0$
          \ForEach {$(V_{\rm{gs}_j},V_{\rm{ds}_j}) \in \bm{V}$}
          \State $I_{\rm sim1} = f(\bm{p},(V_{\rm{gs}_j},V_{\rm{ds}_j}))$
          \State $e=e+(I^{\rm meas}_j - I_{\rm sim1})^2$
          \ForEach {$p_i \in \bm{p}$}
          \State $\bm{p'} = \bm{p}$
          \State Substitute $p'_i = p_i + \delta_i$ for $i$-th element of $\bm{p'}$
            \State $I_{\rm sim2} = f(\bm{p'},(V_{\rm{gs}_j},V_{\rm{ds}_j}))$
            \State $e_{\rm delta}=e_{\rm delta}+(I^{\rm meas}_j - I_{\rm sim2})^2$
          \EndFor
          \State $E= \sqrt{\frac{e}{m}}$
          \State $E_{\rm delta}= \sqrt{\frac{e_{\rm delta}}{m}}$
          \State $\frac{\partial E}{\partial p_i} = \frac{E_{\rm delta}-E}{\delta_i}$
        \EndFor
        \State Return $\nabla E, E$
      \EndFunction
    \end{algorithmic}
  \end{algorithm}
\end{figure}

ND in the context of parameter extraction is computationally intensive
as it involves the calculation of the cost function and gradient with
respect to the model parameters, and hence, the evaluation of the
model equation for all combinations of bias voltages and
parameters. In the ND-based method, $f(\cdot)$ is evaluated
$(1+n)mN_{\rm iter}$ times, where $N_{\rm iter}$ is the iteration
count. The calculation time increases linearly as $n$
increases. Hence, the time required to evaluate the partial derivative
will increase for complex models having larger number of
parameters. Moreover, in the situations where any of the model
parameters are involved in different model equations, the computation
becomes more complex as the parameter extraction has to take all the
model equations into account.  For example, oxide thickness is
involved in both current and capacitance equations in typical MOSFET
models.  The two model equations should be evaluated simultaneously
during the extraction such that the consistent parameter values for
these equations are obtained.  In the optimization, {\em
  gradient\_calc} is performed with respect to each model
equation. Thus, the total calculation complexity is written as the sum
of $(1+n)mN_{\rm iter}$ in each model equation. The proposed method
adopts AD to eliminate the iteration over $n$.

\section{Parameter extraction based on automatic differentiation}
\label{sec:propose}

The proposed novel method of parameter extraction essentially follows the
same procedure as gradient-based extraction shown in
Algorithm~\ref{alg:gd}. By replacing the most time-consuming step of the
differentiation in ND, AD is expected to significantly reduce the time
required for parameter extraction.

\subsection{Automatic Differentiation}
The basic concept of AD\footnote[1]{According to the direction of the
  traverse in the computational graph, the AD is classified into two
  distinct types: forward type and reverse
  type~\cite{JMLR2018_Baydin}. We hereafter call the reverse-type AD
  as AD.}  is the decomposition of partial derivatives using the chain
rule. The derivative, ${\rm d}y/{\rm d}x$, of a composite function, $y
= p(q(x)) = p(w)$, is written using the chain rule as follows:
\begin{equation}
 \frac{{\rm d}y}{{\rm d}x}=\frac{{\rm d}y}{{\rm d}w}\frac{{\rm d}w}{{\rm d}x}.
 \label{eq:example}
\end{equation}
The first and second factors of~(\ref{eq:example}) are individually
calculated because ${\rm d}y/{\rm d}w={\rm d}p(w)/{\rm d}w$ and
${\rm d}w/{\rm d}x={\rm d}q(x)/{\rm d}x$. Then, ${\rm d}y/{\rm d}x$ is
derived as the product of these two factors. In AD, the given expression is
generally represented using a directed acyclic graph
called {\em computational graph}~\cite{NIPS2015_Schulman}.

The AD computes the gradient with respect to each parameter as
the contribution to the output.  This is realized by two calculation
modes: forward and backward.  In the forward mode, the given expression
is first decomposed into a set of primitive expressions (i.e., the
simplest functions) such as addition and multiplication functions; in
this mode, the model parameters and input values are propagated to
obtain the value of the output.  Then, in the backward mode, the output
variable to be differentiated is given first and the partial derivative
value concerning each partial expression (e.g., ${\rm d}y/{\rm d}w$ and
${\rm d}w/{\rm d}x$ in the above example in (\ref{eq:example})) is
recursively calculated. In the gradient calculation involved in AD, the
partial derivatives with respect to each of the input parameters are
obtained simultaneously through a traversal of the computational graph
in the forward and backward directions; hence, there is no need to
repeat the model evaluation for each parameter.

\subsection{Parameter Extraction}

The AD-based gradient calculation, which is executed by the function
{\em gradient\_calc} in line 3 of Algorithm~\ref{alg:gd}, is outlined
in Algorithm~\ref{alg:ad}. Notably, in contrast to
Algorithm~\ref{alg:nd}, there is no nested loop iterating over all
model parameters.  Instead, the forward propagation and backward
propagation are performed in lines 4 and 5, respectively.  $\nabla E$
is derived by summing the gradient for each bias condition, $\nabla
E_{\rm tmp}$.
The input arguments are also simplified than those in ND.  The input
vector $\bm{\delta}$, which represents the small parameter deviation
used for numerically calculating each gradient, is unnecessary in this
method and thus omitted.

\begin{figure}[t!]
  \begin{algorithm}[H]
    \caption{AD-based gradient calculation}
    \label{alg:ad}
    \begin{algorithmic}[1]
      \Function{AD}{$\bm{p}$, ${\bm V}$, $\bm{I}^{\rm meas}$}
      \State Initialize $e$ and $\nabla E$ to zero
      \ForEach {$(V_{\rm{gs}_j},V_{\rm{ds}_j}) \in \bm{V}$}
      \State Calculate $I_{\rm sim} = f(\bm{p},(V_{\rm{gs}_j},V_{\rm{ds}_j}))$ and $e=e+(I^{\rm meas}_j - I_{\rm sim})^2$ through forward mode
      \State Calculate $\nabla E_{\rm tmp}$ and $\nabla E = \nabla E + \nabla E_{\rm tmp}$  through backward mode
      \EndFor
      \State $E= \sqrt{\frac{e}{m}}$
      \State Return $\nabla E, E$
      \EndFunction
    \end{algorithmic}
  \end{algorithm}
\end{figure}

The computational complexity of the forward and backward propagation
is roughly equal to that of a single model
evaluation~\cite{OMS1992_Griewank}; therefore, the calculation
complexity becomes $2mN_{\rm iter}$ and is independent of the number
of parameters $n$, unlike that in ND. Specifically, comparing the
loops in Algorithms~\ref{alg:nd} and~\ref{alg:ad}, the gradient
calculation is accelerated by the factor of $(n+1)/2$ in AD as
compared to ND. Therefore, as the number of the model parameters
increases, the acceleration efficiency linearly improves.  In
situations where two model equations $f_1(\cdot)$ and $f_2(\cdot)$ are
considered, those calculation complexity becomes merely the summation
of that of each equation, $2m_1N_{\rm iter} + 2m_2N_{\rm iter}$. In
the gradient calculation, if $\nabla E$ has the partial derivatives of
the common parameters in the two models, they are added up.

The proposed parameter-extraction method is described using a simple example
model: the drain current equation. The drain current equation is expressed
as $f(\cdot)$ in Algorithm~\ref{alg:ad} and is defined as
follows~\cite{TPEL2018_Shintani}:
\begin{align}
  I_{\rm sim} = \cfrac{1}{1 + \prmt{THETA}\cdot V_{\rm gs}} \, (1 +
  \prmt{LAMBDA}\cdot V_{\rm ds}) \nonumber \\
  \cdot{\prmt{SCALE}}\cdot I_{\rm DD}, \label{eq:equation}
\end{align}
where $I_{\rm sim}$ is the simulated drain current, $I_{\rm DD}$ is an
intermediate value expressed as a function of $V_{\rm gs}$, $V_{\rm ds}$,
and the surface potentials at the metal-oxide-semiconductor (MOS)
interface. The channel current equation includes the channel length
modulation and mobility degradation. $\bf{SCALE}$, $\bf{LAMBDA}$, and
${\bf THETA}$ are model parameters that represent the scaling factor,
channel length modulation, and channel mobility degradation, respectively.

\begin{figure}[!t]
 \centering
 \includegraphics[width=0.52\linewidth]{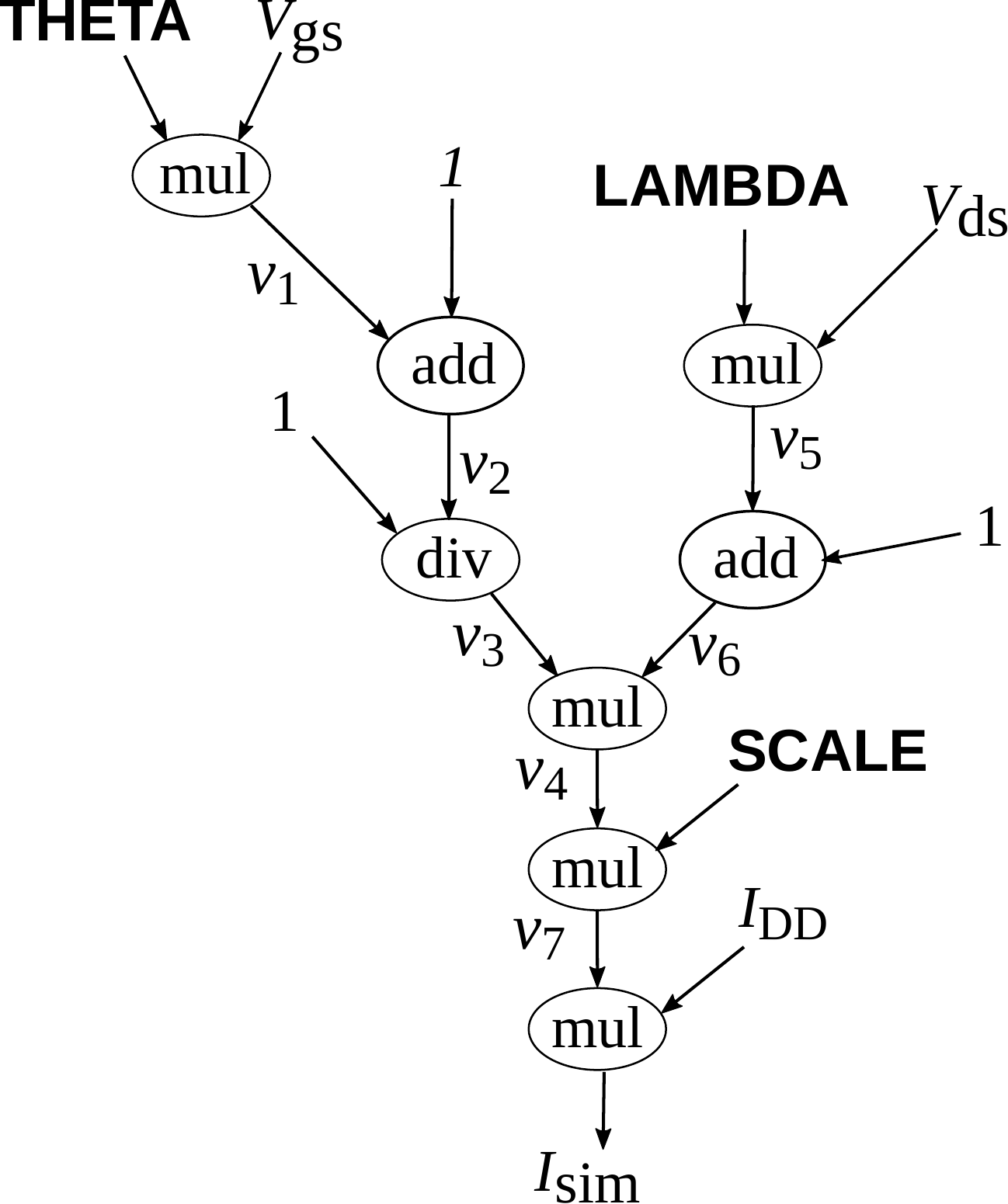}
 \caption{Computational graph of~(\ref{eq:equation}) including five
 multiplication operations, two addition operations, and one division
 operation.}
 \label{fig:graph}
\end{figure}

Fig.~\ref{fig:graph} shows a computational graph
representing~(\ref{eq:equation}). The variables located at the leaves of the
graph (i.e., {\bf THETA} and $V_{\rm ds}$) represent the inputs and that on
the bottom (i.e., $I_{\rm sim}$) represent the output. The vertices labeled
as $v_1, \ldots, v_7$ stand for intermediate variables. The nodes represent
the basic mathematical function, with each node defining an input variable,
output variable, internal connection, and functional behavior. The order of
the input edges is defined consistently to handle the operations that is
commutative such as subtraction and division.

\begin{figure}[!t]
  \centering
  \includegraphics[width=0.73\linewidth]{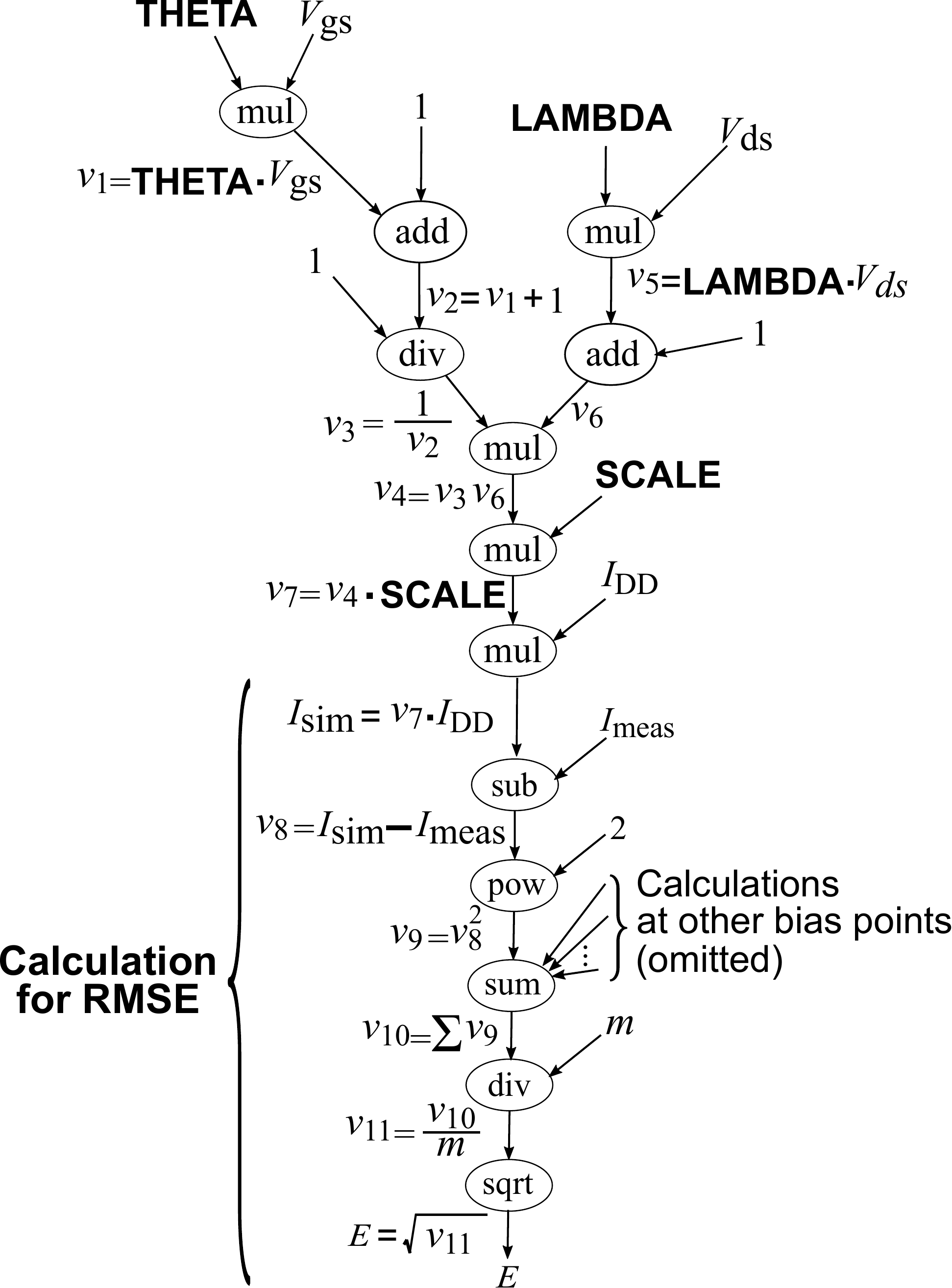}
  \caption{Forward propagation mode. In this figure, the graph is
  simplified by showing only one subgraph for the current calculation
  of $v_9$.}
 \label{fig:forward_graph}
\end{figure}

In the AD-based gradient calculation, the forward propagation is conducted
first. The current model equation is evaluated by traversing the
computational graph to calculate the internal variables under a particular
bias voltage condition, $V_{\rm ds}$ and $V_{\rm gs}$, as shown in
Fig.~\ref{fig:forward_graph}. During the forward propagation, the values of
the internal vertices, $v_1, \ldots, v_{12}$ and $I_{\rm sim}$, are stored
for use in the backward propagation that will be conducted later. Note that,
in Fig.~\ref{fig:forward_graph}, the calculation of the RMSE is in the last
part of the graph. In the summation operation used to calculate the RMSE,
the values of $I_{\rm meas}-I_{\rm sim}$ for each of the bias voltages are
required. To simplify the illustration, only the graph for $v_9$ is shown
and the subgraphs for the other bias points are omitted.

\begin{figure}[!t]
 \centering
 \includegraphics[width=0.73\linewidth]{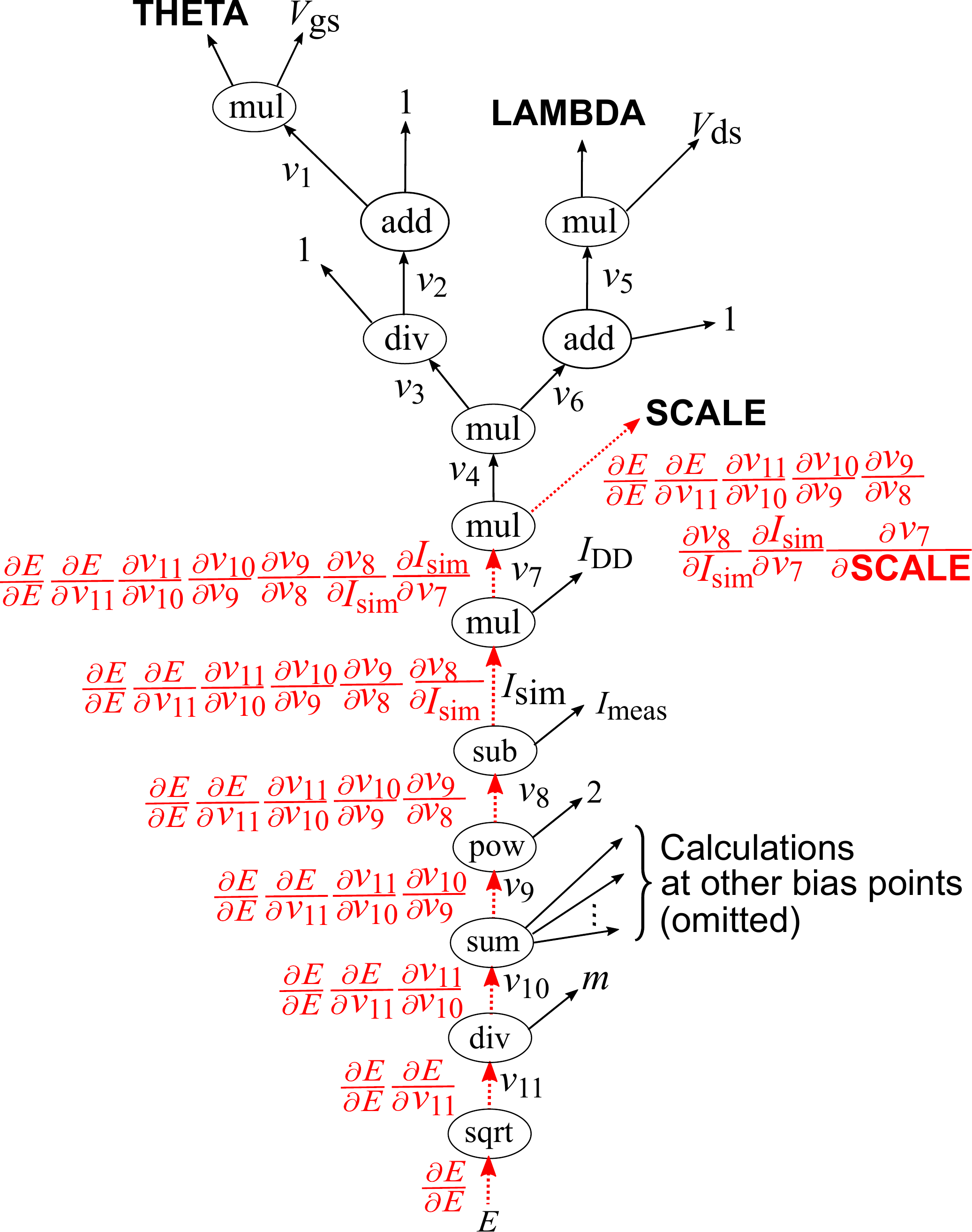}
 \caption{Backward propagation mode.  The directions of arrows are
   reversed as compared to those in Fig.~\ref{fig:forward_graph}.  The
   dashed arrows indicate the path from $E$ to {\bf SCALE}. The
   propagated values on the path in the backward propagation are
   highlighted.}
 \label{fig:backward_graph}
\end{figure}

\begin{figure}[!t]
  \centering
  \subfigure[Multiplication ($x \cdot y$)\label{fig:mul}]{
    \includegraphics[width=0.3\linewidth]{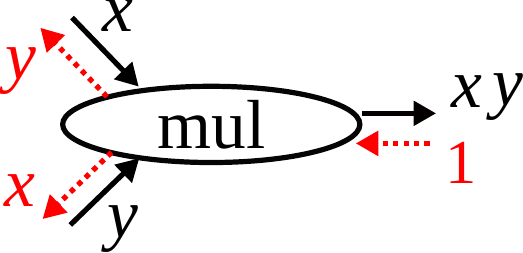}}
  \hspace{2mm}
  \subfigure[Addition ($x + y$)\label{fig:add}]{
    \includegraphics[width=0.3\linewidth]{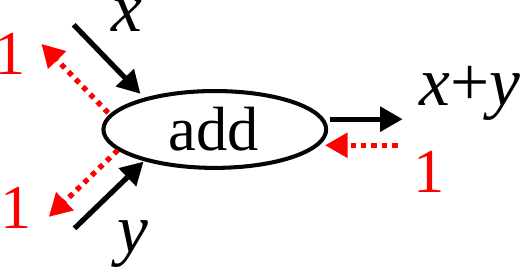}}
  \\
  \subfigure[Division ($x/y$)\label{fig:div}]{
    \includegraphics[width=0.32\linewidth]{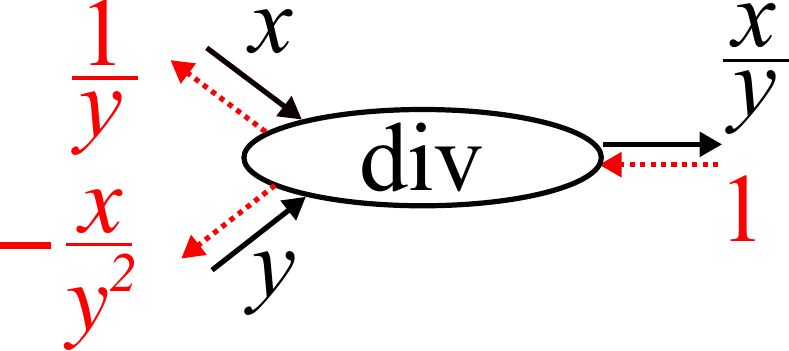}}
  \hspace{2mm}
  \subfigure[Summation ($\sum x_m$)\label{fig:sum}]{
    \includegraphics[width=0.32\linewidth]{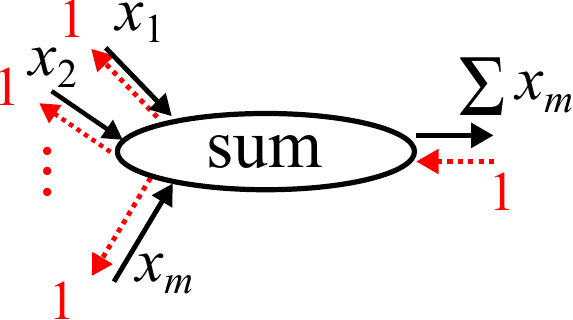}}
  \\
  \subfigure[Exponential ($\exp(x)$)\label{fig:exp}]{
    \includegraphics[width=0.4\linewidth]{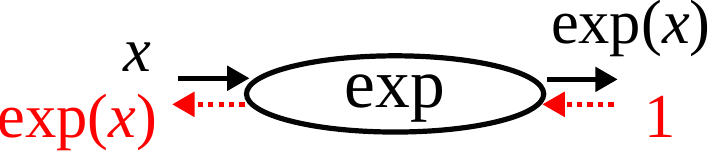}}
  \\
  \hspace{2mm}
 \subfigure[Square root ($\sqrt{x}$)\label{fig:sqrt}]{
 \includegraphics[width=0.32\linewidth]{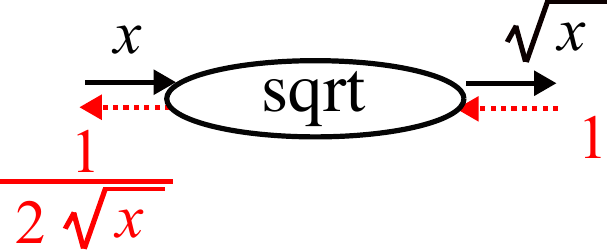}}
\caption{Representative forward and backward propagation rules.  The solid
  arrow is for forward propagation and the dashed one is for the backward
  propagation.  In this figure, a value of 1 is assumed to be the input in
  the backpropagation.} \label{fig:rules}
\end{figure}

In the backward mode, the derivative of the output $E$ is propagated
backwards through the graph to calculate the partial derivative with
respect to each model parameter, as shown in
Fig.~\ref{fig:backward_graph}, in which the path to calculate
$\frac{\partial E}{\partial {\bf{SCALE}}}$ has been highlighted as an
example.  There are eight vertices ($E$, $v_{11}$, $v_{10}$, $v_9$,
$v_8$, $I_{\rm sim}$, $v_7$, and ${\bf{SCALE}}$) on the path.  According
to the chain rule, $\frac{\partial E}{\partial {\bf{SCALE}}}$ is
calculated as follows:
\begin{eqnarray}
 \label{eqchain}
  \frac{\partial E}{\partial {\bf{SCALE}}} =
  \frac{\partial E}{\partial v_{11}} \cdot \frac{\partial v_{11}}{\partial v_{10}}
  \cdot \frac{\partial v_{10}}{\partial v_9} \cdot \frac{\partial v_9}{\partial v_8}
  \nonumber  \\
  \cdot \frac{\partial v_8}{\partial I_{\rm sim}}
  \cdot \frac{\partial I_{\rm sim}}{\partial v_7}
  \cdot \frac{\partial v_7}{\partial {\bf{SCALE}}}.
\end{eqnarray}
The derivative of each edge with respect to the previous edge can be
easily obtained: $\frac{\partial E}{\partial
v_{11}}=\frac{1}{2\sqrt{v_{11}}}$ because $E=\sqrt{v_{11}}$.  Similarly,
$\frac{\partial I_{\rm sim}}{\partial v_7}=I_{\rm DD}$ as $I_{\rm
sim}=v_7\cdot I_{\rm DD}$, and $\frac{\partial v_7}{\partial
{\bf{SCALE}}}=v_4$ as $v_7=v_4 \cdot {\bf{SCALE}}$.  Substituting all
the derivatives in~(\ref{eqchain}) with the partial derivatives of
the edges obtained by traversing the graph, the following equation is
derived:
\begin{equation}
  \frac{\partial E}{\partial {\bf{SCALE}}}= \frac{1}{2\sqrt{v_{11}}} \cdot
  \frac{1}{m} \cdot 2v_8 \cdot I_{\rm DD} \cdot v_4.
  \label{eq:lambda}
\end{equation}
The values of $v_{11}$, $v_8$, and $v_7$ were already stored while
progressing the forward mode.  The partial derivatives are calculated
for all model parameters and the resulting set of partial derivatives is
used as a gradient in the parameter updating phase. The chain rule
provides us with a way to easily and efficiently calculate the gradients
without approximation.
%
The edges of the graphs for forward and backward propagations can be
determined by simply consulting the rules as summarized in
Fig~\ref{fig:rules}.

We would like to note that the partial derivative shown
in~(\ref{eq:lambda}) is the value of a single bias condition.  When
there are multiple bias conditions, which is usually the case, the
backward propagation explained above is carried out for all bias
conditions.  That is, at the sum node, the partial difference of all
the paths contributing $E$ is calculated for each bias condition until
the model parameter node is reached.  The partial derivatives of the
model parameters are the summation of the partial derivatives of the
different bias voltages diverged at the sum node.

The computational graph must be constructed only once for a given
MOSFET model. The construction can be very quick as described in the
experimental section. Additionally, once the computational graph is
generated, it can be reused as long as the same MOSFET model is used;
therefore, the time needed to construct the computational graph is
virtually negligible. Besides, since building a computational graph on
the basis of model equations, either model equations or source code,
at least, is required as a prerequisite for applying the proposed
parameter extraction.  Note that the extracted model parameters by the
proposed method bears physical meaning when physics-based model is
used as the computational graph is consistent with the model
equations.
Thus, the estimated model parameters can be used for device
characterization, such as manufacturing variability analysis.

\section{Experiments}\label{sec:exp}

To quantitatively evaluate the effectiveness of the proposed method, the
model parameters for two different MOSFET models were extracted using AD-
and ND-based parameter extraction. A commercially available silicon carbide
(SiC) power MOSFET~\cite{sct2450ke} was used for these measurements.
The experiments were conducted using a Linux PC with an Intel Xeon W5590
3.33\,GHz central processing unit (CPU) using a single thread. The
extraction was implemented using the Python programming language.

\begin{figure}[!t]
  \centering
  \includegraphics[width=0.38\linewidth]{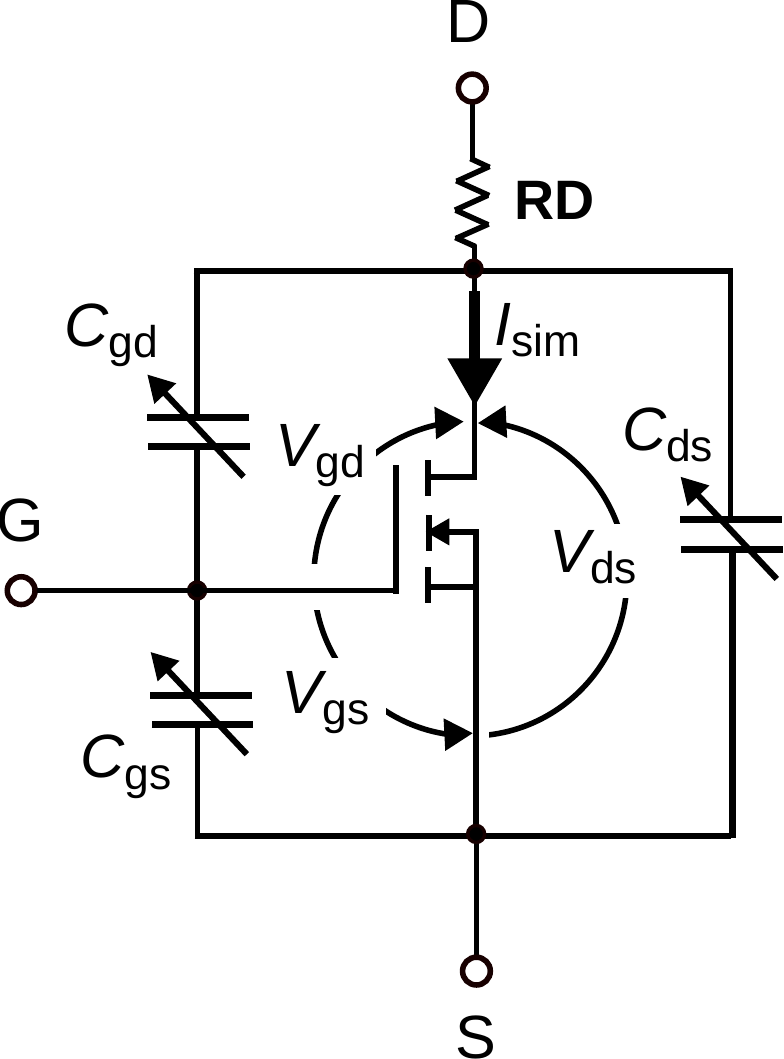}
  \caption{Equivalent circuit of SiC power MOSFET.}
  \label{fig:equiv}
\end{figure}

\subsection{MOSFET Models}\label{sec:model}

The two MOSFET models used in this section are an $N$-th-power-law
model~\cite{TED1991_sakurai} and a surface-potential-based
model~\cite{TPEL2018_Shintani}.  The equivalent circuit of SiC power
MOSFET is shown in Fig.~\ref{fig:equiv}.  The MOSFET model
characterizes the current characteristics $I_{\rm sim}$ and three
terminal capacitances: gate-source capacitance $C_{\rm gs}$,
drain-source capacitance $C_{\rm ds}$, and gate-drain capacitance
$C_{\rm gd}$.  The parameter extraction for the $N$-the-power-law
model uses $I_{\rm sim}$ only, while that for the
surface-potential-based model uses $I_{\rm sim}$, $\cds$, and $\cgd$
in this experiment. Note that $\cgs$
is modeled as constant as in~\cite{TPEL2018_Shintani}.


\subsubsection{$N$-th-Power-Law Model}

In this model, the drain current is calculated based on the threshold
voltage, $\prmt{VTH}$, of the MOSFET\@. The saturation voltage,
$V_\mathrm{ds,sat}$, and saturation current, $I_\mathrm{d,sat}$, are
defined as
\begin{align}
  V_\mathrm{ds,sat}&=\prmt{J}\left(V_\mathrm{gs}-\prmt{VTH}\right)^{\prmt{M}}
                     \,\,\,\,\,\textrm{and} \\
  I_\mathrm{d,sat}&=\prmt{K}\left(V_\mathrm{gs}-\prmt{VTH}\right)^{\prmt{N}},
\end{align}
where $\prmt{J}$ and $\prmt{M}$ are the fitting parameters used to
calculate the current in the linear region and $\prmt{K}$ and $\prmt{N}$
are other fitting parameters that are used for the saturation
region. $V_\mathrm{ds,mod}$ replaces $V_\mathrm{ds}$ to represent a
smooth transition between the linear and saturation
regions~\cite{TPEL2018_Shintani}:
\begin{align}
  V_\mathrm{ds,mod} = \frac{V_\mathrm{ds}}{\left[1+\left(\frac{V\mathrm{ds}}
  {V_\mathrm{ds,sat}}\right)^{\prmt{DELTA}}
  \right]^{\frac{1}{\prmt{DELTA}}}},
  \label{eq:delta}
\end{align}
where $\prmt{DELTA}$ controls the smoothness of the transition between
$V_\mathrm{ds,mod}$ and $V_\mathrm{ds,sat}$. The drain current,
$I_\mathrm{sim}$, is calculated based on the channel length modulation
and mobility degradation~\cite{TED2006_Miura,TED2006_Gildenblat} as
follows:
\begin{align}
  I_\mathrm{sim}=&I_\mathrm{d,sat}\left(2-\dfrac{V_\mathrm{ds,mod}}{V_\mathrm{ds,sat}}\right)
   \dfrac{V_\mathrm{ds,mod}}{V_\mathrm{ds,sat}}\times\nonumber\\
   &(1+\prmt{LAMBDA} \cdot V_\mathrm{ds})\,\left[1+\prmt{THETA}\,(V_\mathrm{gs}-\prmt{VTH})\right].
\end{align}
The model parameters of the $N$-th-power-law model are listed in
Table~\ref{tab:sakurai_parameters}.

\begin{table}[t]
  \centering
  \caption{Model parameters of the $N$-th-power-law model~\cite{TED1991_sakurai}}
    \label{tab:sakurai_parameters}
    \begin{tabular}{l|l}\hline
      Parameter & Description \\ \hline
      {\bf VTH} & Threshold voltage [V]  \\
      {\bf K} & Fitting parameter for the linear region [-]  \\
      {\bf M} & Fitting parameter for the linear region [-]\\ 
      {\bf J} & Fitting parameter for the saturation region [-]  \\
      {\bf N} & Fitting parameter for the saturation region [-]  \\
      {\bf LAMBDA} & Channel length modulation [1/V]  \\
      {\bf THETA} & Mobility degradation [1/V]  \\
      \prmt{DELTA}& Smoothing parameter for gradual transition\\
                & between the linear and saturation regions [-]\\ \hline
      \end{tabular}
\end{table}

\subsubsection{Surface-Potential-Based Model}

\begin{table}[!t]
  \centering
  \caption{Model parameters of the current equation in the surface-potential-based model~\cite{TPEL2018_Shintani}}
    \label{tab:sp_parameters}
    \begin{tabular}{l|l}\hline
      Model parameter & Description \\ \hline
      \prmt{TOX} &Oxide thickness [m]\\
      \prmt{VFBC}&Flat-band voltage of the channel region [V]\\
      \prmt{NA}& Acceptor concentration [$\rm cm^{-3}$]\\
      \prmt{SCALE}& Current gain factor [$\rm cm^2/V$]\\
      \prmt{RD}& Parasitic resistance at the drain side [$\Omega$]\\
      \prmt{LAMBDA}& Channel length modulation [1/V]\\
      \prmt{THETA}& Channel mobility degradation [1/V]\\
      \prmt{DELTA}& Smoothing parameter for gradual transition\\
                & between the linear and saturation regions [-]\\
      \hline
    \end{tabular}
\end{table}

\begin{table}[!t]
  \centering
  \caption{Model parameters of the capacitance equation in the surface-potential-based model~\cite{TPEL2018_Shintani}}
  \label{tab:symbol_cv}
  \begin{tabular}{l|l}\hline
    Model parameter & Description \\ \hline
    \prmt{ADS}& Drain-source overlap area [$\rm cm^2$]\\
    \prmt{ND}&Donor concentration [$\rm cm^{-3}$]\\
    \prmt{VBI}& Built-in potential of PN junction [V]\\
    \prmt{COXD}& Gate-drain oxide capacitance [F]\\
    \prmt{AGD}& Gate-drain overlap area [$\rm cm^2$]\\
    \prmt{VFBD}& Gate-drain flat-band voltage [V]\\ \hline
  \end{tabular}
\end{table}

\begin{table}[!t]
  \centering
  \caption{Physical constants}
  \label{tab:physical_parameters}
  \begin{tabular}{l|l|r}\hline
    Parameter & Description & Value \\ \hline
    $k$ & Boltzmann's constant [$\rm J/K$] & $1.38 \times 10^{-23}$\\
    $q$ & Elementary charge [C] &$1.60\times 10^{-19}$\\
    $T$& Absolute temperature [K] & 298\\
    $\phi_{\rm t}$& Thermal voltage ($kT/q$) [V] & 0.026\\
    $\varepsilon_{\rm SiC}$ &Permittivity of SiC [F/m] & $9.7\times 8.85 \times 10^{-12}$\\
    $\varepsilon_\mathrm{ox}$ & Permittivity of gate oxide [F/m]& $3.9\times8.85\times10^{-12}$\\
    $n_i$ & Intrinsic carrier concentration & $4.82\times10^{15}$\\ 
     & of SiC [$\rm cm^{-3}$] & \\
    \hline
  \end{tabular}
\end{table}
The proposed parameter extraction is also performed on the
surface-potential-based model, which was developed to simulate the
behavior of a SiC power MOSFET~\cite{TPEL2018_Shintani}. The model
parameters of the current and capacitance model equations are listed
in Tables~\ref{tab:sp_parameters} and~\ref{tab:symbol_cv},
respectively.  The physical constants for these models are summarized
in Table~\ref{tab:physical_parameters}.

According to the current model in~\cite{TPEL2018_Shintani}, the surface
potentials, $\phi_\mathrm{sS}$, and $\phi_\mathrm{sD}$ are first
calculated for the source and drain ends of the channel as functions of
$V_{\rm gs}$ and $V_{\rm ds}$. The inverted charge of the channel is
determined as a function of the surface potential. Then, the
intermediate value $I_{\rm DD}$ in~(\ref{eq:equation}) is computed based
on $\phi_{\rm sS}$ and $\phi_{\rm sD}$ as follows:
\begin{align}
  I_{\rm DD}&=C_{\rm ox}{  (V_{\rm gs} -
        {\bf VFBC} + \phi_{\rm t})(\phi_{\rm sD} - \phi_{\rm sS})} \nonumber \\
  &- \cfrac{1}{2}\, C_{\rm ox}
    (\phi_{\rm sD}^2 - \phi_{\rm sS}^2) \nonumber \\
  &-\cfrac{2}{3}\, \phi_{\rm t} \gamma \left\{\left(\phi_{\rm sD}/\phi_{\rm t}
        - 1\right)^{\frac{3}{2}} - (\phi_{\rm sS}/\phi_{\rm t} -
    1)^{\frac{3}{2}}
    \right\} \nonumber \\
  &+\phi_{\rm t} \gamma
      \left\{\left(\phi_{\rm sD}/\phi_{\rm t} - 1\right)^{\frac{1}{2}} -
    (\phi_{\rm
    sS}/\phi_{\rm
    t} - 1 )^{\frac{1}{2}}\right\},
    \label{eq:idd}
\end{align}
where
\begin{align}
\gamma&=\sqrt{2\varepsilon_{\rm SiC}kT\cdot {\bf NA}}.
\end{align}
Here, $k$ and $T$ are the Boltzmann's constant and the absolute temperature,
respectively. $\varepsilon_\mathrm{SiC}$ is the permittivity of SiC,
$\phi_{\rm t}$ is the thermal voltage, and $C_{\rm ox}$ is gate oxide
capacitance per unit area. Further,
$C_{\rm ox}=\varepsilon_{\rm ox}/{\bf TOX}$, where $\varepsilon_{\rm ox}$
and {\bf TOX} are the permittivity and thickness, respectively, of the gate
oxide. {\bf VFBC} and {\bf NA} are the flat-band voltage and the acceptor
concentration, respectively. The channel current model also includes a
smooth transition between the linear and saturation regions according to the
model parameter {\bf DELTA}, which is defined by a similar equation to that
in the $N$-th-power-law model. Finally, substituting $I_{\rm DD}$
in~(\ref{eq:idd}), the simulated drain current, $I_{\rm sim}$, is
derived. The drain current in the structure also depends on the parasitic
resistance {\bf RD}~\cite{TED2013_Mattausch} which reduces internal drain
voltage of the MOSFET.

In the surface-potential-based model, the calculation of $\phi_{\rm sS}$
and $\phi_{\rm sD}$ involves solving a non-linear
equation~\cite{Baliga_book}, whose solution cannot be explicitly
expressed. Typically, the surface potentials have to be obtained using
iterative methods~\cite{TED2006_Miura,TED2013_Mattausch} such as the
Newton-Raphson method. Though the computational graph
may also be constructed for the model equations
containing
iterations,
we adopt the following closed-form expression for the surface potential
$\phi_{\rm s}$~\cite{SSE2001_Chen} for simplifying the
calculation graph:
\begin{align}
  (V_{\rm gs}-\prmt{VFBC}-\phi_{\rm s})^2= \gamma^2 \phi_{\rm t}
  [(\exp(-\frac{\phi_{\rm s}}{\phi_{\rm t}})+\frac{\phi_{\rm s}}{\phi_{\rm t}}-1) +
  \nonumber \\
  \exp(-(2\phi_{\rm F} + \phi_{\rm t})/\phi_t)(\exp(\frac{\phi_{\rm s}}{\phi_{\rm
  t}}) - \frac{\phi_{\rm s}}{\phi_{\rm t}} -1)].
  \label{eq:spe}
\end{align}
$\phi_{\rm sS}$ and $\phi_{\rm sD}$ are derived by solving~(\ref{eq:spe})
with respect to~$\phi_{\rm s}$ at $\phi_{\rm F}=0$ and
$\phi_{\rm F}=V_{\rm ds}$, respectively. Thus, the model equations of the
surface-potential-based model can be entirely represented by a computational
graph.

The capacitance characteristics, $\cds$ and $\cgd$, are expressed in the
surface-potential-based model as follows:
\begin{align}
  C_{\rm ds} &= {\bf ADS}\cdot
               \sqrt{\frac{q \cdot \varepsilon_{\rm SiC}\cdot {\bf ND}}
               {2({\bf VBI} + V_{\rm ds}) }}\,\,\,\,\,\,{\rm and}
               \label{eq:cds} \\
  C_{\rm gd} &= {\bf  COXD} \parallel C_{\rm dep}.
               \label{eq:cgd}
\end{align}
$\cds$ is a bias-dependent junction capacitance between drain and
source, which is calculated on the basis of the capacitance model of
the PN junction as shown in~(\ref{eq:cds}).  $\cgd$ is modeled as a
series connection of the constant gate oxide capacitance ${\bf COXD}$
and the bias-dependent depletion-layer MOS capacitance $C_{\rm dep}$.
Here, $C_{\rm dep}$ is a MOS capacitance represented as a function of
the surface potential $\phi_{\rm gd}$ of the channel formed on the
drain region under the junction field effect transistor (JFET)
region. It can be written as shown in~(\ref{eq:cdep}), where
$\phi_{\rm gd}$ is computed in a similar way to the calculation of
$\phi_{\rm s}$ using \eqref{eq:spe} with ($V_{\rm gd}$, $V_{\rm ds}$,
${\bf ND}$, ${\bf VFBD}$).

\begin{figure*}[!t]
  \begin{equation}
      C_{\rm dep} = {\bf AGD}\cdot\sqrt{2q\varepsilon_{\rm SiC}\cdot{\bf
          ND}}
      \, \cfrac{1 - e^{-\phi_{\rm gd}/\phi_{\rm t}} + e^{-(2\phi_{\rm F}+
          V_{\rm ds})/
          \phi_{\rm t}} (e^{\phi_{\rm gd}/\phi_{\rm t}} - 1) }
      {2\sqrt{\phi_{\rm t}e^{-\phi_{\rm gd}/\phi_{\rm t}} + \phi_{\rm gd}
          - \phi_{\rm t} + e^{-(2\phi_{\rm F}+V_{\rm ds})/\phi_{\rm t}}
          (\phi_{\rm t}e^{\phi_{\rm gd}/\phi_{\rm t}} - \phi_{\rm gd} - \phi_{\rm t}) }}
      \label{eq:cdep}
    \end{equation}
\end{figure*}

\subsubsection{Initial Parameter Determination}\label{sec:initial}

\begin{figure}[!t]
  \begin{minipage}{0.48\hsize}
    \begin{center}
      \includegraphics[width=.9\columnwidth]{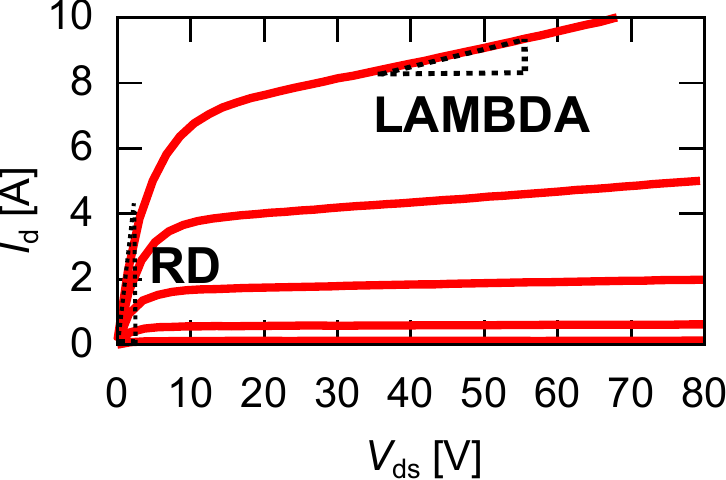}
    \end{center}
    \caption{Extraction of {\bf LAMBDA} and {\bf RD}.}
    \label{fig:ext_idvd}
  \end{minipage}
  \hspace{2mm}
  \begin{minipage}{0.48\hsize}
    \begin{center}
      \includegraphics[width=.98\columnwidth]{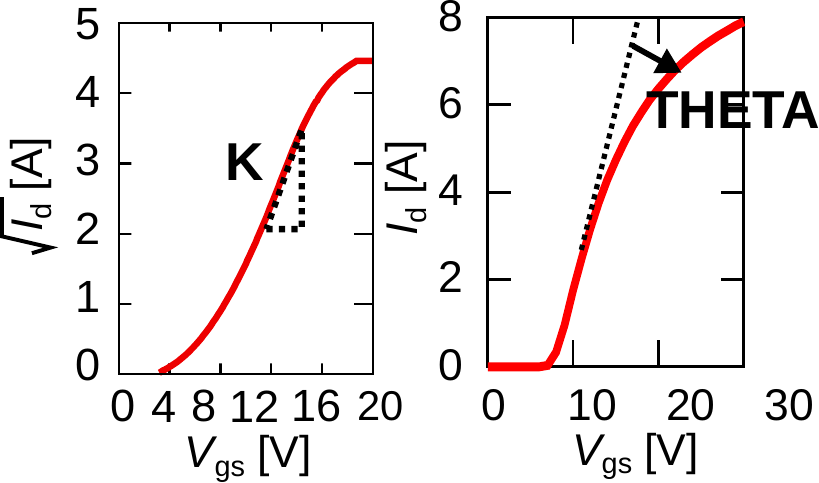}
    \end{center}
    \caption{Extraction of {\bf K} and {\bf THETA}.}
    \label{fig:ext_idvg}
  \end{minipage}
  \hspace{2mm}
  \begin{minipage}{0.48\hsize}
    \vspace{2mm}
    \begin{center}
      \includegraphics[width=.98\columnwidth]{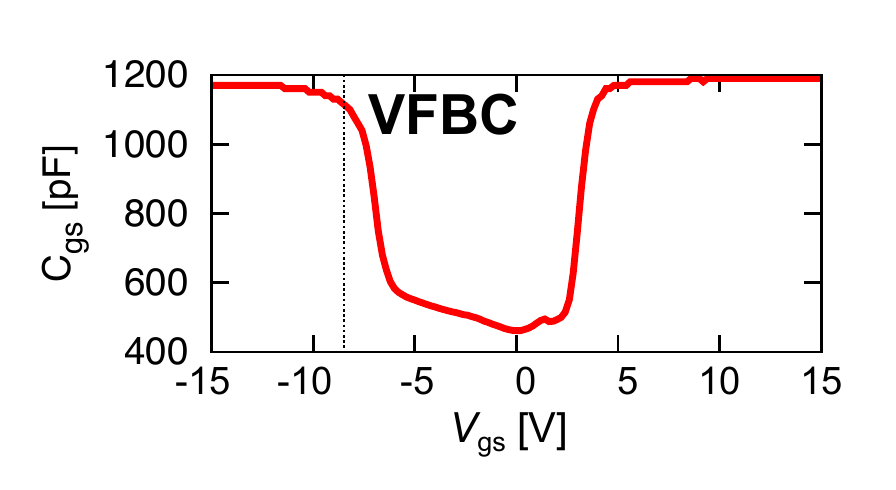}
    \end{center}
    \caption{Extraction of {\bf VFBC}.}
    \label{fig:ext_vfbc}
  \end{minipage}
  \hspace{2mm}
  \begin{minipage}{0.48\hsize}
    \begin{center}
      \includegraphics[width=.9\columnwidth]{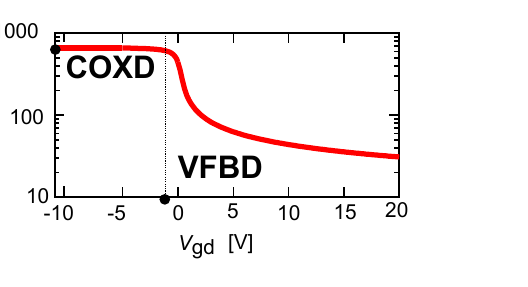}
    \end{center}
    \caption{Extraction of {\bf COXD} and {\bf VFBD}.}
    \label{fig:coxd_vfbd}
  \end{minipage}
\end{figure}

The initial parameter determination is one of the important steps in
the model development because the model parameters are repeatedly
updated to minimize the discrepancy with the measured device
characteristics, starting from the initial values.  When the choice of
the initial parameters is inappropriate, the model parameters may be
determined as a local optimum solution that can be far from the ground
truth. The standardized compact models provide the initial parameter
determination as well as the model
descriptions~\cite{BSIM48,HiSIM-HV}.  In the experiment, we apply an
initial parameter determination for the surface-potential-based model
proposed in~\cite{WIPDA2019-Shintani}.  In this work, we assume {\bf
  TOX} is 50\,nm. The default value of {\bf DELTA}, which is
introduced to model the gradual transition between the linear and
saturation regions, is set to 0.8~\cite{HiSIM-HV}.

The slope of the $I_{\rm d}$-$V_{\rm ds}$ curve in the saturation
region is represented by {\bf LAMBDA}, while that at $V_{\rm ds}
\simeq 0$\,V of a high $V_{\rm gs}$ shows {\bf RD} as shown in
Fig.~\ref{fig:ext_idvd}.  {\bf K} and {\bf THETA} are extracted from
the $I_{\rm d}$-$V_{\rm gs}$ curve as shown in
Fig.~\ref{fig:ext_idvg}.  {\bf K} is approximated by the slope of
saturation-region current, and {\bf THETA} is extracted from the
linear region current.  {\bf VFBC} is the flat-band voltage of the MOS
interface at the channel region.  Beyond that voltage, the depletion
capacitance becomes apparent in the MOS capacitance characteristics.
Hence, {\bf VFBC} can be estimated from the gate-source capacitance
$C_{\rm gs}$ against $V_{\rm gs}$ at which $C_{\rm gs}$ starts to
bend, as shown in Fig.~\ref{fig:ext_vfbc}.

As a good estimate, we approximate $C_{\rm dep}$ to a PN
junction capacitance that depends on drain-source voltage
$\vgd$~\cite{EPE2011_Phankong}:
\begin{eqnarray}
  C_{\rm dep} = {\bf AGD}
  \sqrt{\frac{q \cdot \varepsilon_{\rm SiC}\cdot {\bf ND}}{2({\bf VFBD}-\vgd)}}.
  \label{eq:approximated_cgd}
\end{eqnarray}
As shown in Fig.~\ref{fig:coxd_vfbd}, {\bf VFBD} and {\bf COXD} are
obtained as the capacitance in the accumulation mode. {\bf AGD} is estimated
as $\frac{\bf TOX}{\varepsilon_{\rm SiC}{\bf COXD}}$. Then, by substituting
{\bf VFBD}, {\bf COXD}, and {\bf AGD} to~(\ref{eq:approximated_cgd}), {\bf ND}
can be calculated. Finally, {\bf NA} is obtained from
\begin{equation}
 {\bf VBI} = \frac{kt}{q}\ln \left(\frac{{\bf NA} \cdot {\bf ND}}{n_i^2}
 \right),
 \label{eq:vbi}
\end{equation}
where $n_i$ is the intrinsic carrier concentration. {\bf VBI} is derived
as the forward voltage where the body-diode current starts to flow.
{\bf ADS} is calculated by substituting {\bf VBI} and {\bf ND} to (\ref{eq:cds}).

\subsubsection{Computational Graph}
\label{sec:graph}

\begin{figure}[!t]
  \centering
  \includegraphics[width=1.0\linewidth]{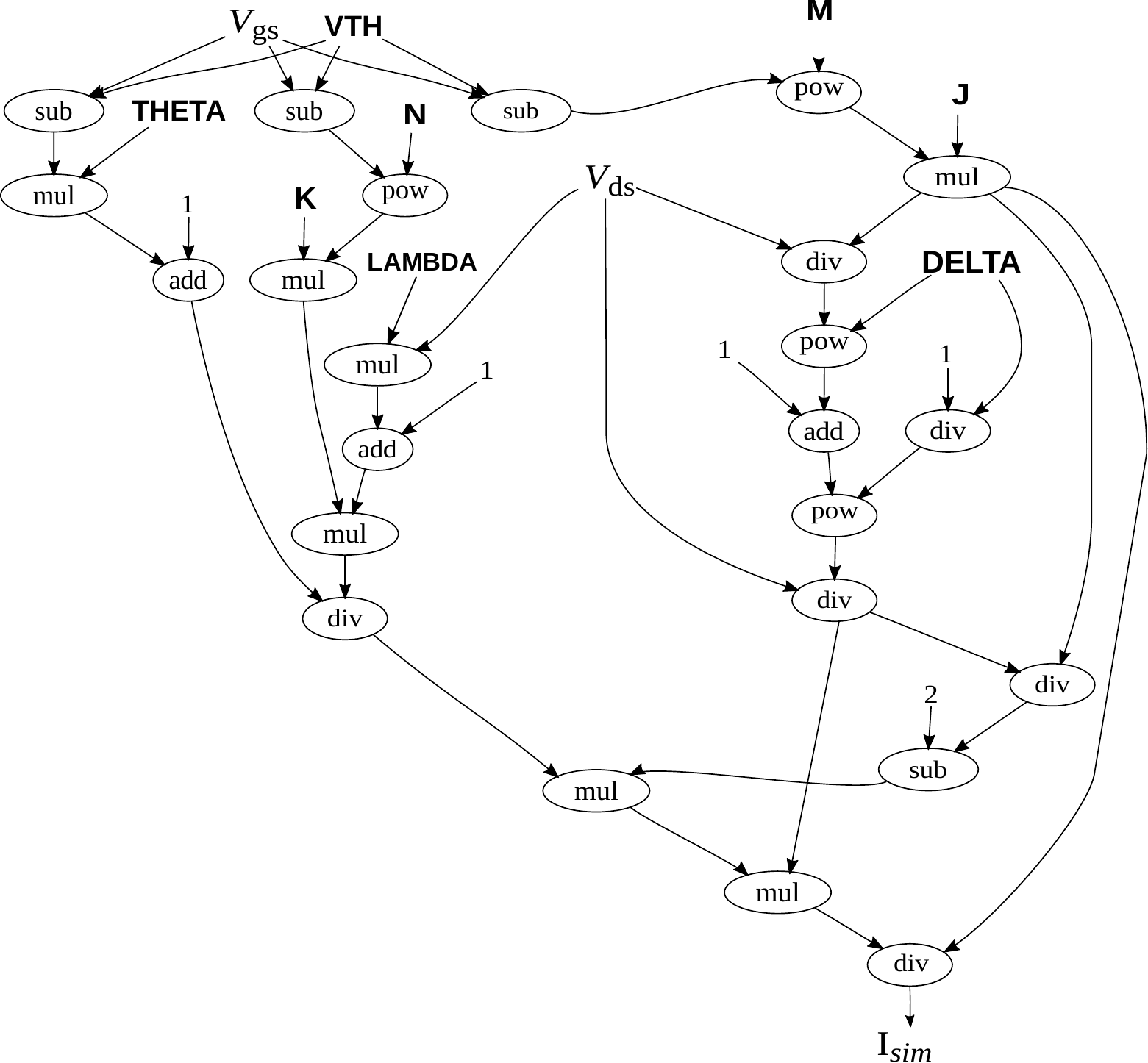}
  \caption{Computational graph of the $N$-th-power-law model for a particular bias voltage.}
  \label{fig:graph_n-th-power}
\end{figure}

The computational graphs of the two models were constructed based on
the respective model equations.  Graph manipulations are implemented
using a Python package, NetworkX~\cite{SCIPY2008_Hagberg}.  As shown
in Table~\ref{tab:generation_time}, the execution time was much less
than 0.1\,second on the single thread PC for both models.
Table~\ref{tab:graph_size} summarizes the sizes of the two
computational graphs from the input parameters to $I_{\rm sim}$ for a
particular bias voltage.  The subgraph for calculating $I_{\rm sim}$
is reused for different bias voltages. The graph size of the
surface-potential-based model is approximately ten times larger than
that of the $N$-th-power-law model. Thus, only the computational graph
of the $N$-th-power-law model is presented in
Fig.~\ref{fig:graph_n-th-power}.  Through defining the forward and
backward functions of each node and incorporating the computational
network into Algorihgm~\ref{alg:ad}, the AD-based parameter extraction
is carried out according to Algorithm~\ref{alg:gd}.

\begin{table}[!t]
  \caption{Graph generation time}\label{tab:generation_time}
  \centering
  \begin{tabular}{l|l||r}
    \hline
    \multicolumn{2}{l||}{Model}    & Time {[}s{]} \\ \hline
    \multicolumn{2}{l||}{$N$-th-power-law model} & 0.0384         \\ \hline
    \multirow{3}{*}{Surface-potential-based model}  & $I_{\rm sim}$ & 0.0571 \\ \cline{2-3}
                                   & $\cds$     & 0.0039        \\ \cline{2-3}
                                   & $\cgd$     & 0.0288         \\ \hline
  \end{tabular}
\end{table}

\begin{table}[!t]
  \caption{Size of the computational graphs}\label{tab:graph_size}
  \centering
  \begin{tabular}{l|l||r|r}
    \hline
    \multicolumn{2}{l||}{Model}    & No. of Edges & No. of Vertices \\ \hline
    \multicolumn{2}{l||}{$N$-th-power-law model} &  39 & 48 \\ \hline
    \multirow{3}{*}{Surface-potential-based model} & $I_{\rm sim}$ & 384 & 540 \\ \cline{2-4}
                                   & $\cds$ & 20 & 20 \\ \cline{2-4}
                                   & $\cgd$ & 318 & 228 \\ \hline
  \end{tabular}
\end{table}


\subsection{Simulation Setup}
I-V curves of the SiC MOSFET~\cite{sct2450ke} were measured at room
temperature using a dedicated curve tracer~\cite{ICMTS2016_Nakamura}
while sweeping $V_{\rm gs}$ from 6\,V to 14\,V with 2\,V steps and
sweeping $V_{\rm ds}$ from 0\,V to 50\,V with 2\,V steps (the total
number of bias voltages tested $m$, were 125). C-V curves were
measured by a commercial curve tracer~\cite{b1505a} at 1\,MHz. The
data points are 300 for both $\cds$ and $\cgd$ measurements.  The
parameter updating process was conducted by using
AdaGrad~\cite{JMLR2011_Duchi}:
\begin{eqnarray}
  h_i &=& h_i + \left(\frac{\partial E}{\partial
          p_i}\right)^2 \qquad {\mathrm{and}}\label{eq:adagrad1} \\
  p_i &=& p_i - \eta_i \frac{1}{\sqrt{h_i}} \frac{\partial E}{\partial p_i}.
  \label{eq:adagrad2}
\end{eqnarray}
The parameter update is performed to decrease the RMSE between the
measured drain current and that obtained through the MOSFET model with
the latest model parameters.  In AdaGrad, the parameter vector,
$\bm{h}=(h_0,...,h_{n-1})$, is introduced. Based on the gradient,
$\eta_i \frac{1}{\sqrt{h_i}}$ is gradually reduced as shown
in~(\ref{eq:adagrad1}) and~(\ref{eq:adagrad2}). Generally, the initial
values of all elements of $\bm{h}$ were set to zero and all elements of
$\bm{\eta}$ were set to 1/100 for the respective parameter values.

\subsection{Results}
\subsubsection{Parameter extraction for current characteristics}\label{sec:c-1}
\begin{figure}[t!]
  \centering
  \subfigure[$N$-th-power-law model \label{fig:npower}]{
    \includegraphics[width=0.55\linewidth]{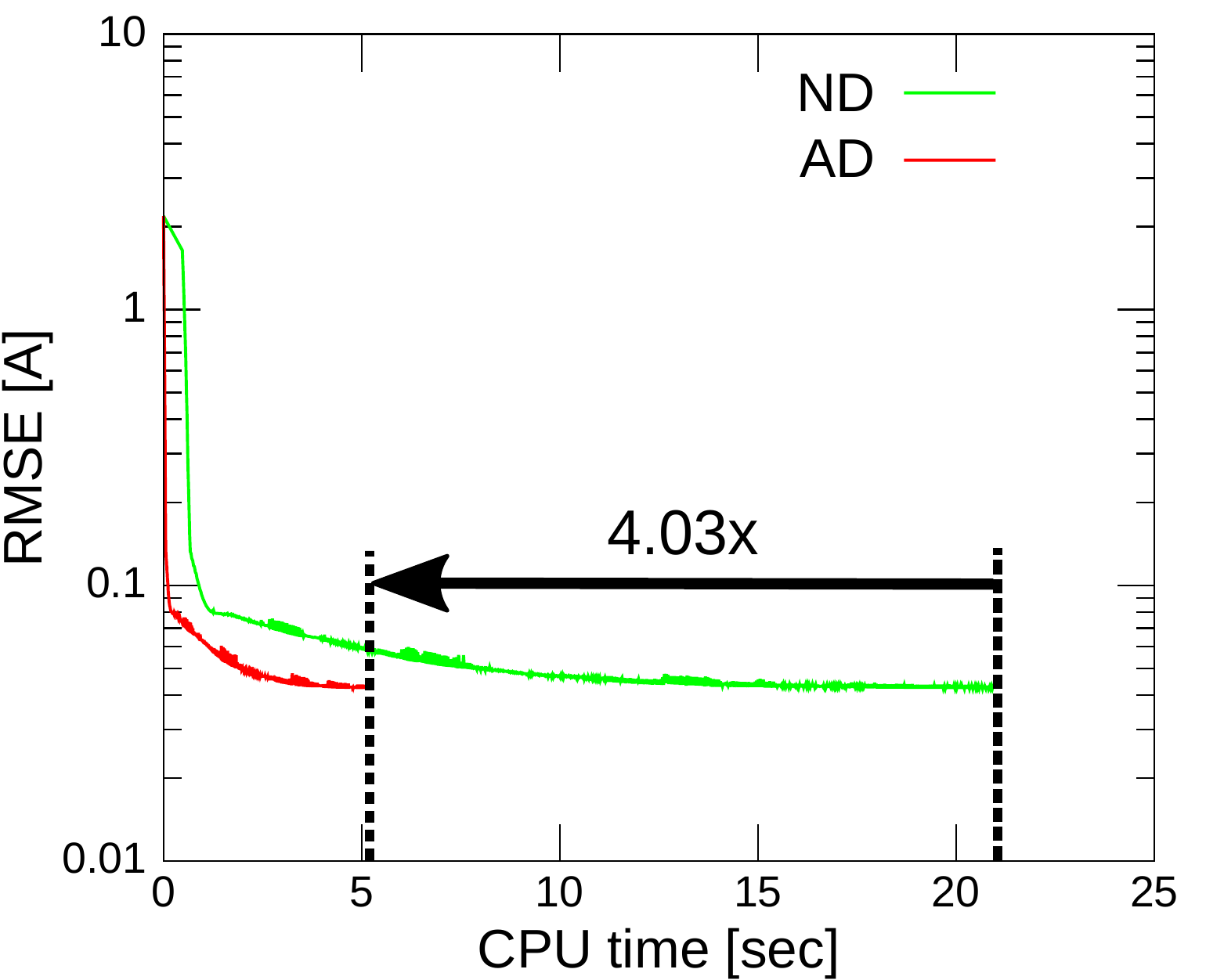}}\\
  \subfigure[Surface-potential-based model\label{fig:sp}]{
    \includegraphics[width=0.55\linewidth]{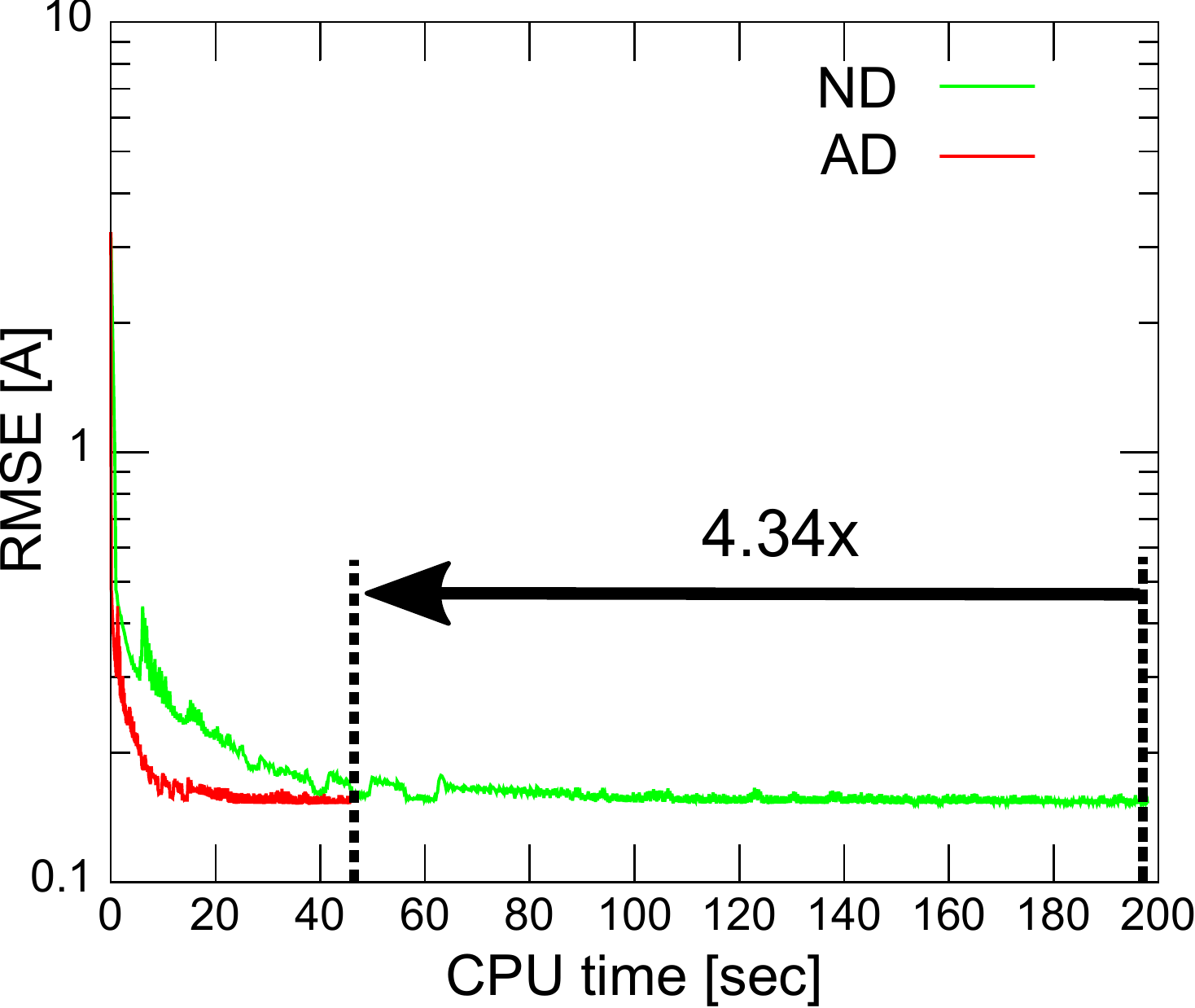}}
  \caption{RMSE as a function of the computation time.} \label{fig:result1}
\end{figure}

The parameters extracted by the proposed AD-based approach was
compared with those by the ND-based method using each of the two
MOSFET models. In this experiment, the maximum number of iterations,
$N_{\rm max}$, was set to 1,000 for both the models.  Also, $E_{\rm
  target}$ was set to 0.04\,A and 0.16\,A for each model.  The initial
values are randomly determined for the $N$-th-power-law model, while
those of the surface-potential-model are determined by the initial
parameter determination procedure described in
Section~\ref{sec:initial}. Fig.~\ref{fig:result1} shows the RMSE as
function of the computation time.  The proposed method accelerated the
parameter extraction by 4.03$\times$ and 4.34$\times$ for the
$N$-th-power-law model and the surface-potential-model, respectively.
This improvement in the computation time was close to the theoretical
value of 4.5 ($= (8+1)/2$) for a generalized model with eight
parameters.

\begin{figure}[t!]
  \centering
  \subfigure[$N$-th-power-law model \label{fig:sakurai}]{%
    \includegraphics[width=.62\columnwidth]{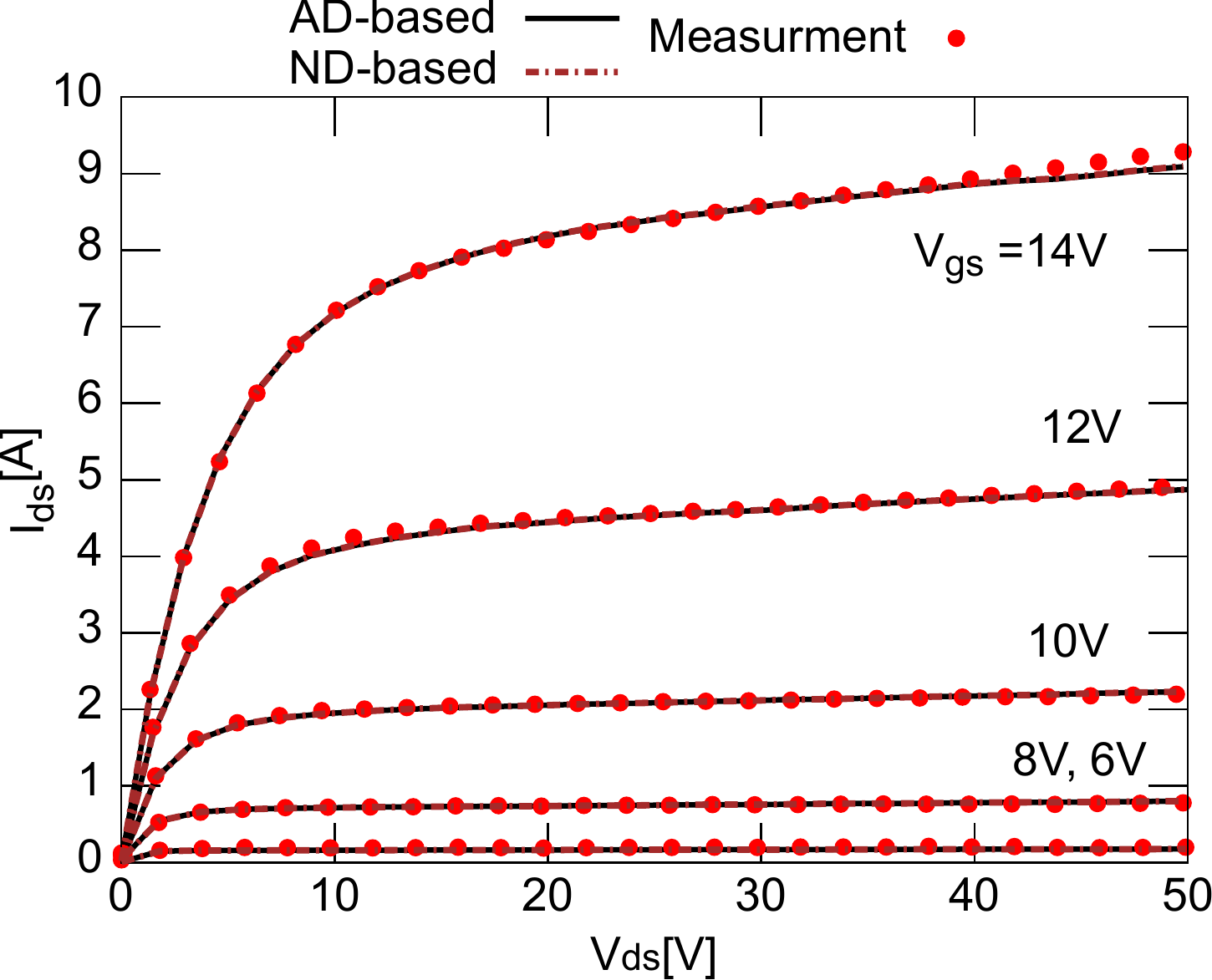}
  }
  \\
  \centering
  \subfigure[Surface-potential-based model \label{fig:iv_sp}]{%
    \includegraphics[width=.65\columnwidth]{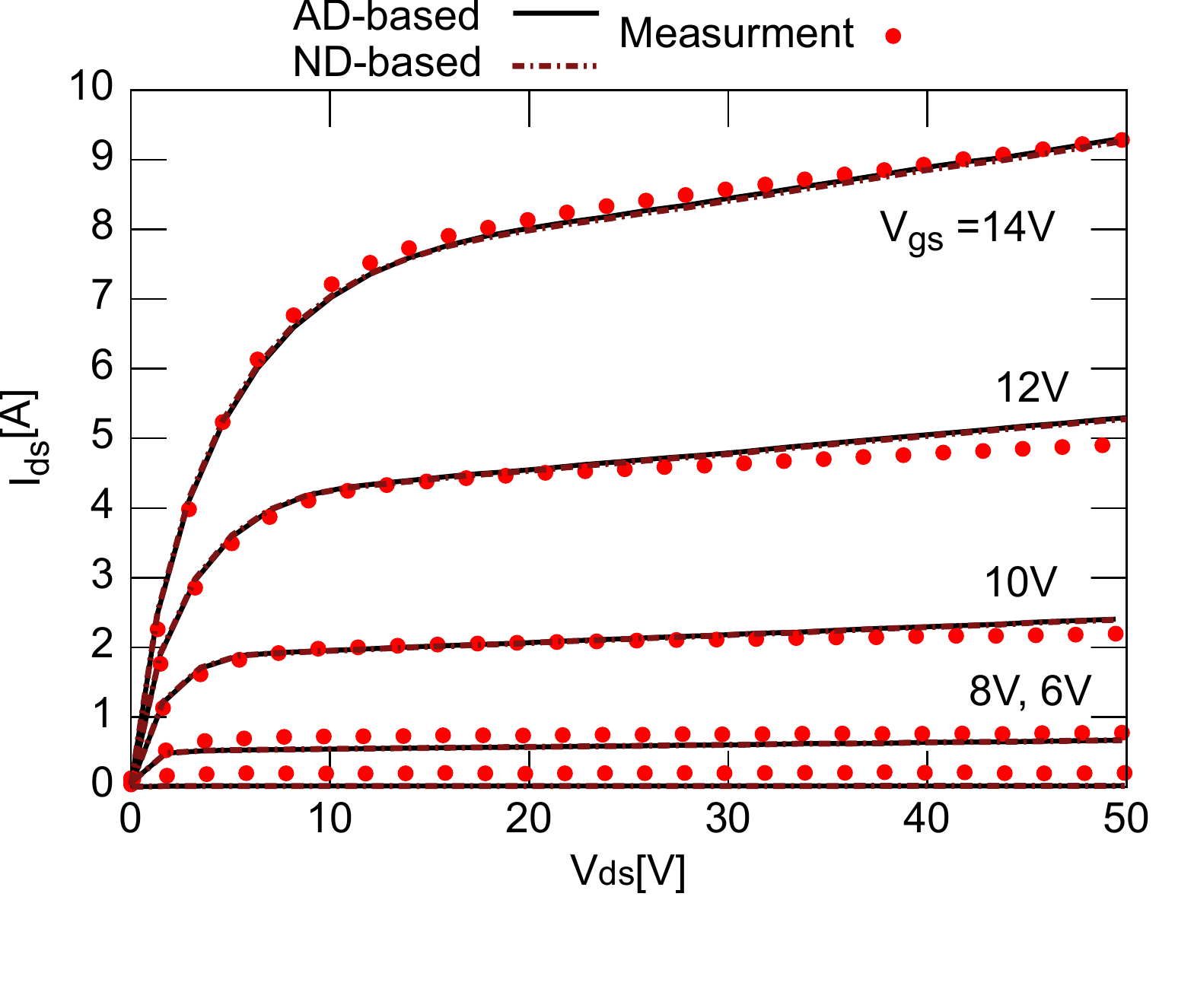}
  }
  \caption{Simulated and measured I-V curves.}
  \label{fig:iv}
\end{figure}

The measured and simulated I-V characteristics of the two models are
presented in Fig.~\ref{fig:iv}. The final model parameters extracted using
the AD and ND methods are used to generate the graph. These results show
that both MOSFET models accurately reproduced the I-V characteristics of the
SiC MOSFET\@. The extracted parameter values at $N=N_{\rm max}$ are
summarized in Tables~\ref{tab:sakurai_error} and~\ref{tab:sp_error}. Also,
the initial values are listed in Table~\ref{tab:sp_error} for the
surface-potential-based model. The relative error was smaller than 2\% in
all cases. In addition, it can be seen from Table~\ref{tab:sp_error} that
the initial and the optimized parameters are close with each other,
suggesting that the good initial parameters have been extracted through the
procedure. Hence, it was concluded that the proposed method
accelerates parameter extraction without affecting the accuracy.

\begin{table}[!t]
  \centering
  \caption{Extracted values for the $N$-th-power-law model}
  \label{tab:sakurai_error}
  \begin{tabular}{l||r|r|r} \hline
    &  &  & Relative  \\
    Parameter & AD & ND &  error [\%]  \\ \hline
    {\bf VTH} & 2.600 & 2.600 & 0.116 \\
    {\bf K} & $ 2.691 \times 10^{-3}$& $2.681 \times 10^{-3}$& 0.372 \\
    {\bf N} & 3.284 & 3.286 & 0.061\\
    {\bf LAMBDA} & $2.606 \times 10^{-3}$ & $2.259\times10^{-3}$ & 0.576 \\
    {\bf THETA} & $3.440 \times 10^{-4}$ & $3.464 \times 10^{-4}$ &0.610 \\
    {\bf M} & 1.743 & 1.744 & 0.057 \\
    {\bf J} &0.119 & 0.119 & 0.000 \\
    {\bf DELTA} & 1.269 & 1.267 & 0.158 \\ \hline
  \end{tabular}
\end{table}

\begin{table}[!t]
  \centering
  \caption{Extracted values for the surface-potential-based model}
  \label{tab:sp_error}
  \begin{tabular}{l||r|r|r} \hline
    Parameter&  &  & Relative  \\
    (Initial value) & AD & ND & error [\%]  \\ \hline
    {\bf SCALE} & & &  \\
    (5166360) & 5403054 & 5398779 & 0.079 \\
    {\bf TOX} & && \\
    ($5.0\times 10^{-8}$) & $4.788\times10^{-08}$ & $4.790\times10^{-08}$ & 0.035 \\
    {\bf NA} & &&\\
    ($1.31\times 10^{17}$) & $1.313\times10^{17}$ &$1.313\times10^{17}$& 0.000 \\
    {\bf LAMBDA} & &&\\
    ($8.69\times 10^{-3}$) & $6.110\times10^{-3}$ & $6.086\times10^{-3}$& 0.387 \\
    {\bf VFBC} & &&\\
    ($-4.90$) & $-1.812\times10^{-3}$ &$-1.780\times10^{-3}$ & 1.869  \\
    {\bf THETA} & &&\\
    ($5.91\times10^{-3}$) & $5.912\times10^{-8}$ &$5.941\times10^{-8}$ & 0.492 \\
    {\bf DELTA} & &&\\
    ($0.80$) & 0.6170 & 0.6150 & 0.329 \\
    {\bf RD} & &&\\ 
    ($2.90\times10^{-3}$) & $2.7178\times10^{-3}$ &$2.737\times10^{-3}$ & 0.715  \\ \hline
  \end{tabular}
\end{table}

\subsubsection{Parameter extraction for multiple
      objectives}\label{sec:multi}
The proposed method can handle model parameters used in different
model equations.  Since {\bf TOX} and {\bf NA} are contained in the
current and capacitance characteristics, $\cds$ and $\cgd$, in the
surface-potential-based model, these parameters have to be consistent
for both characteristics.  In this experiment, the computational
graphs were constructed so that the parameters are simultaneously
optimized for all the model equations.
The cost function is set as the normalized sum of the residuals
to balance the weights of each characteristic.
The initial values of the parameters are determined through the
determination procedure.  In this experiment, $E_{\rm target} = 0.02$
was added as the termination condition, in addition to iteration count
$N_{\rm iter} = 1000$.  Since {\bf VBI} can be represented using {\bf
  NA} and {\bf ND} as shown in (\ref{eq:vbi}), the total number of the
model parameters thus is 13 excluding {\bf VBI}.

\begin{figure}[!t]
  \centering
  \includegraphics[width=0.55\linewidth]{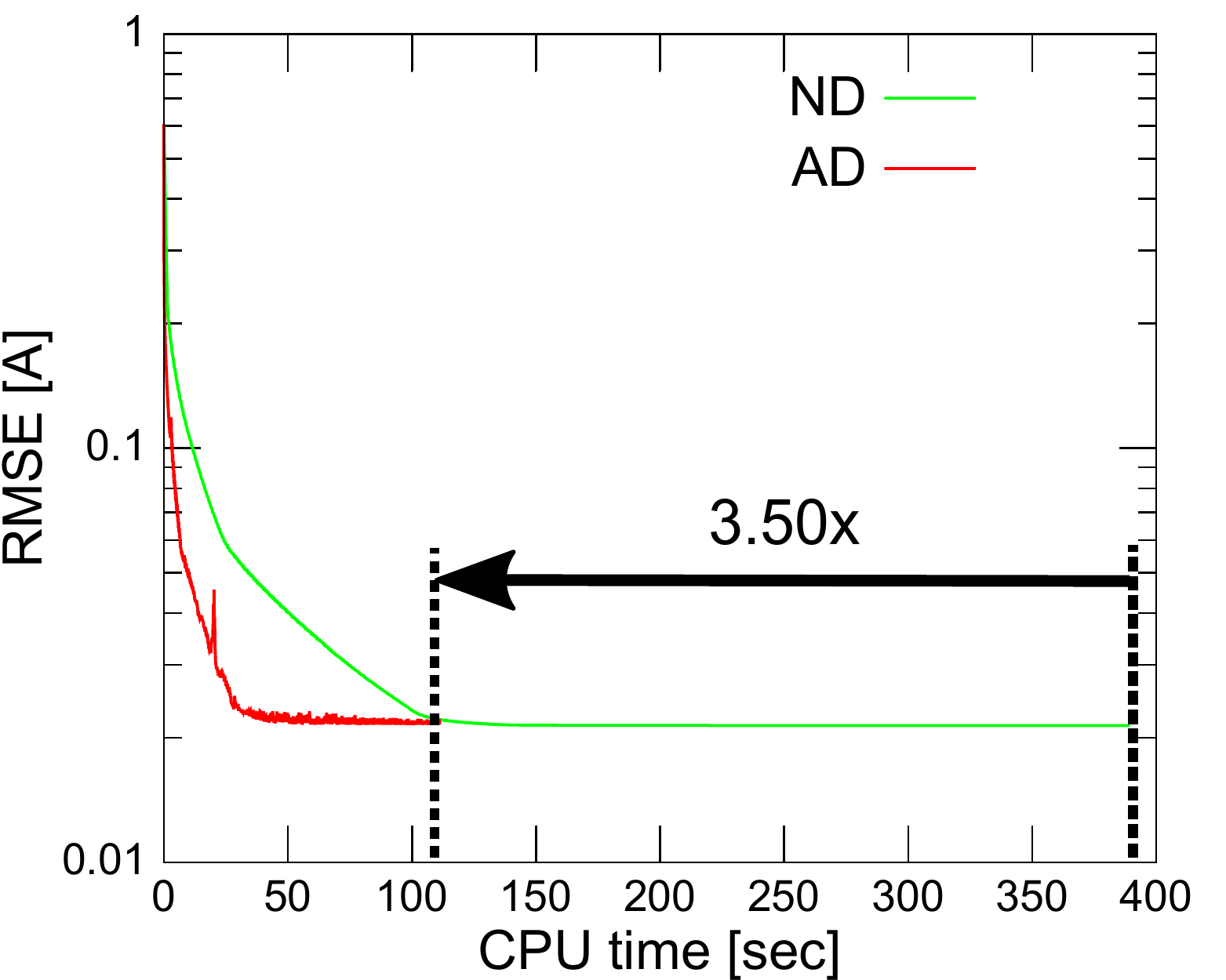}
  \caption{RMSE as a function of the computation time for the
        simultaneous parameter extraction.}
  \label{fig:multi_rmse}
\end{figure}

\begin{figure}[t!]
  \centering
  \subfigure[I-V curve]{%
    \includegraphics[width=.65\columnwidth]{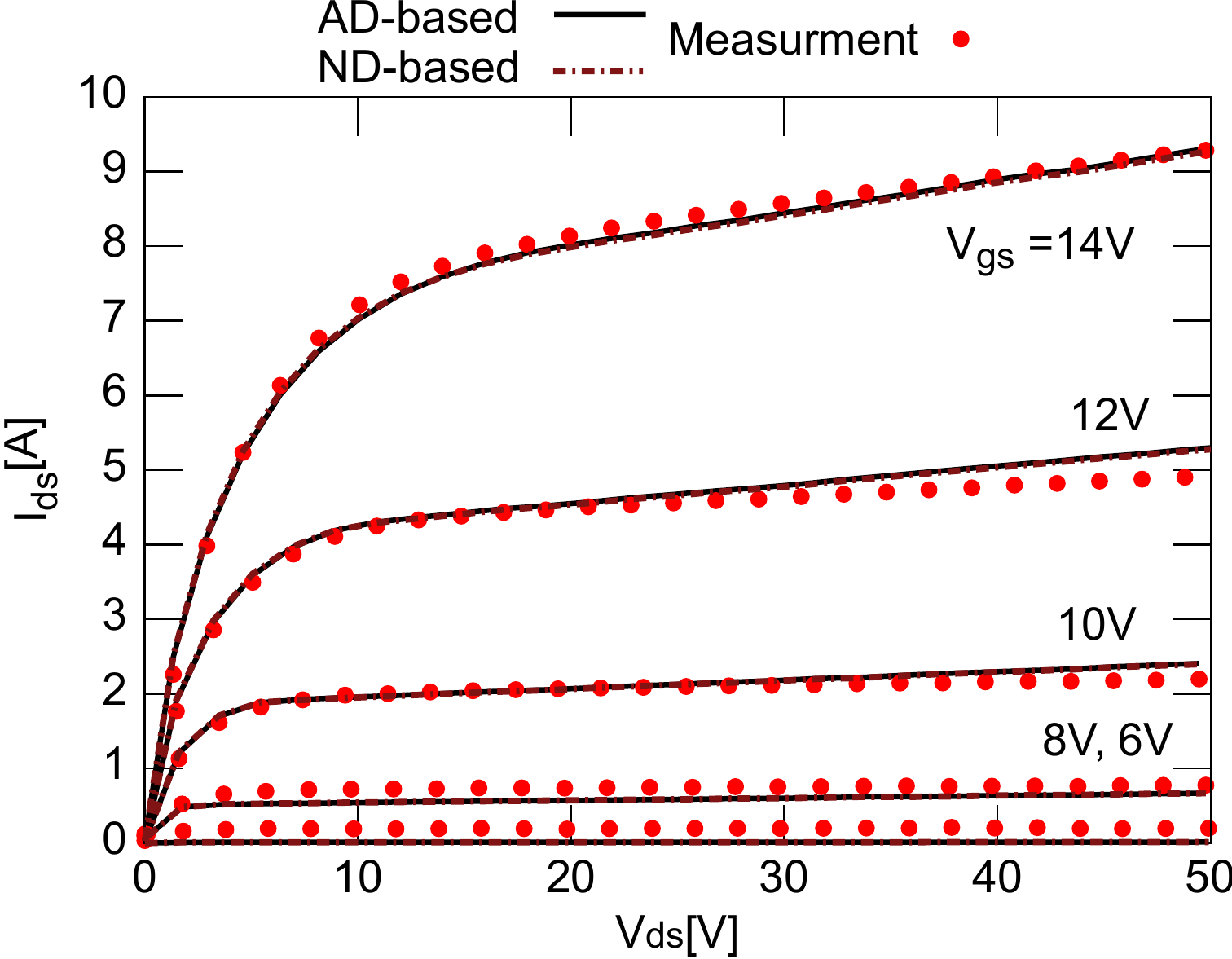}
  }
  \\
  \centering
  \subfigure[C-V curve]{%
    \includegraphics[width=.65\columnwidth]{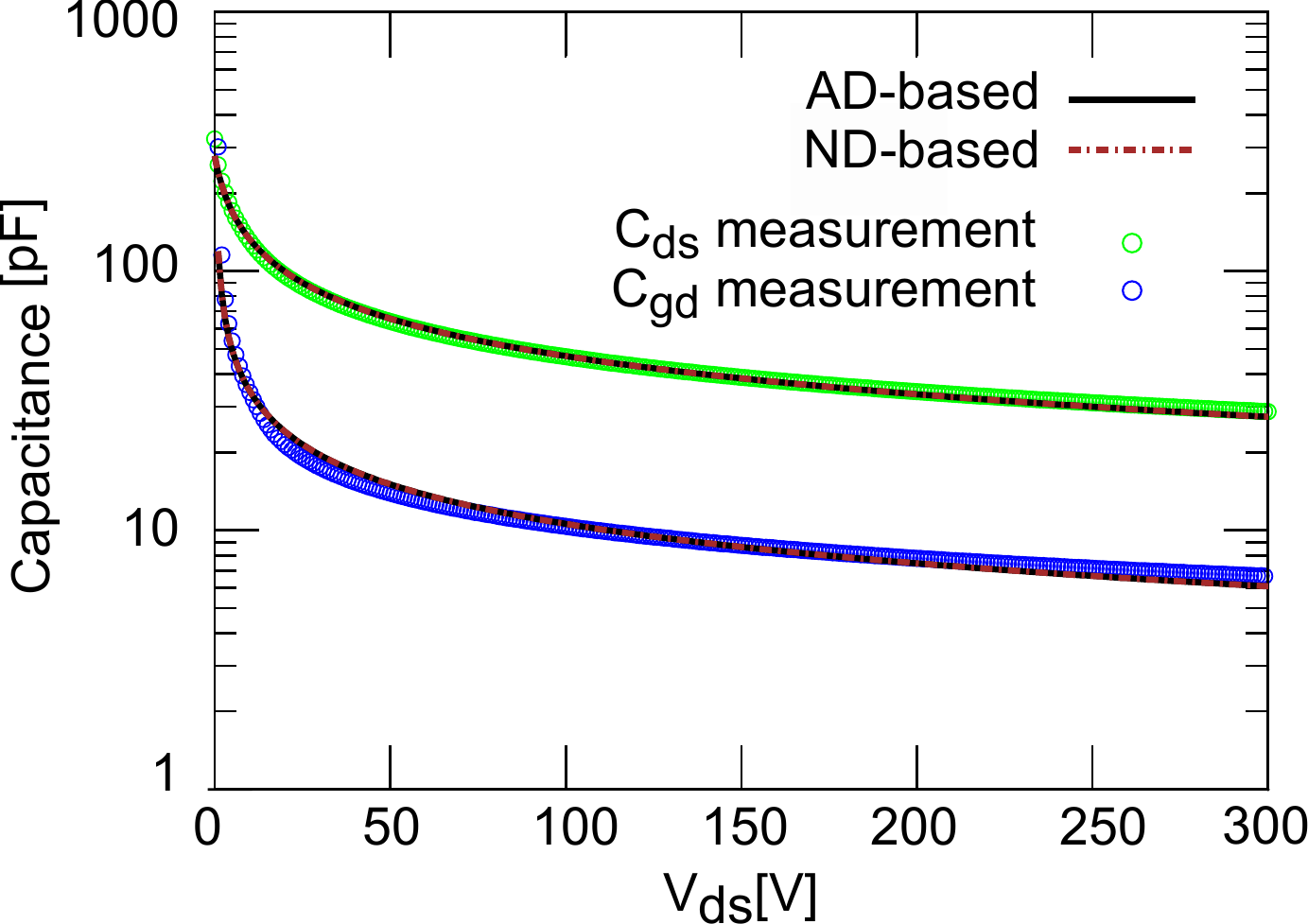}
  }
  \caption{Simulated and measured curves.}
  \label{fig:multi_fit}
\end{figure}

Fig.~\ref{fig:multi_rmse} shows the RMSE as function of computation
time for the simultaneous parameter extraction.  The proposed method
calculates the model parameters 3.50$\times$ faster than the ND-based
parameter extraction. The measured and simulated I-V and C-V
characteristics of the models are presented in
Fig.~\ref{fig:multi_fit}. The graphs are plotted by the final model
parameters extracted using the AD and ND methods. Form the results,
good agreement can be seen among the I-V and C-V characteristics.  The
initial and extracted parameter values when the algorithm exits are
summarized in Table~\ref{tab:sp_multi_error}. The relative error
between the parameters extracted by using the AD and ND methods was
smaller than 3\% in all the parameters. From the result, we confirmed
that the proposed method is applicable for the capacitance
characteristics, and the simultaneous parameter extraction of the
different characteristics.

\begin{table}[!t]
  \centering
          \caption{Extracted values for $I_{\rm sim}$, $\cds$, and $\cgd$ by the
            simultaneous extraction}\label{tab:sp_multi_error}
          \begin{tabular}{l||r|r|r} \hline
            Model parameter & &  & Relative   \\
            (Initial value) & AD & ND & error [\%]  \\ \hline
            {\bf SCALE} &&&\\
            (5166360)  & 5644684 & 5574780 & 1.238\\
            {\bf TOX}  &&& \\
            ($5.00\times10^{-08}$) & $4.933\times10^{-08}$ & $4.947\times10^{-08}$ & 0.286 \\
            {\bf NA} &&& \\
            ($1.31\times10^{17}$) & $1.313\times10^{17}$ & $1.313\times10^{17}$ & 0.000 \\
            {\bf LAMBDA} &&& \\
            ($8.69\times10^{-3}$) & $6.119\times10^{-3}$ & $6.083\times10^{-3}$ & 0.589 \\
            {\bf VFBC} &&&\\
            ($-4.90$) & $-1.943$ & $-1.985$ & 2.168 \\
            {\bf THETA} &&&\\
            ($5.910\times10^{-3}$) & $5.927\times10^{-3}$ & $5.910\times10^{-3}$ & 0.287 \\
            {\bf DELTA} &&&\\
            (0.80) & 0.6073 & 0.6130 & 0.938  \\
            {\bf RD} &&&\\
            ($2.90\times10^{-3}$) & $2.021\times10^{-3}$ & $2.057\times10^{-3}$ & 1.744 \\ \hline
            {\bf ADS} &&&\\
            (0.00776)  & 0.0250 & 0.0250 & 0.000 \\
            {\bf ND}  &&& \\
            ($5.27\times10^{15}$) & $5.266\times10^{15}$ & $5.266\times10^{15}$ & 0.000 \\
            {\bf COXD} &&& \\
            ($4.36\times10^{-10}$) & $4.360\times10^{-10}$ & $4.360\times10^{17}$ & 0.000 \\
            {\bf VFBD} &&& \\
            ($1.00$) & 0.1055 & 0.1055 & 0.000 \\
            {\bf AGD} &&& \\
            ($6.31\times10^{-5}$) & $5.549\times10^{-3}$ & $5.549\times10^{-3}$ & 0.000 \\\hline
          \end{tabular}
\end{table}

\subsubsection{Parameter extraction using LM method}
In addition to the gradient-descent-based method, the proposed method
can be applied to various optimization algorithms that rely on
derivatives.  Here, we show the result of the parameter extraction
derived by incorporating the proposed method with the LM method, which
is adopted in a widely used industrial device modeling
flow~\cite{iccap}.

\begin{figure}[t!]
  \centering
  \subfigure[Number of iterations\label{fig:lm_iter}]{
    \includegraphics[width=0.55\linewidth]{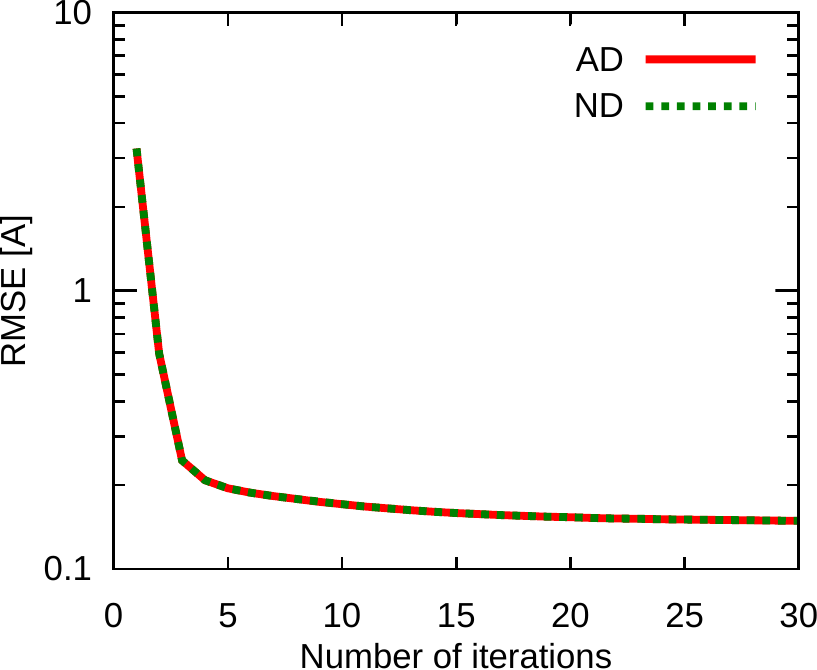}}\\
  \subfigure[CPU time\label{fig:lm_time}]{
    \includegraphics[width=0.55\linewidth]{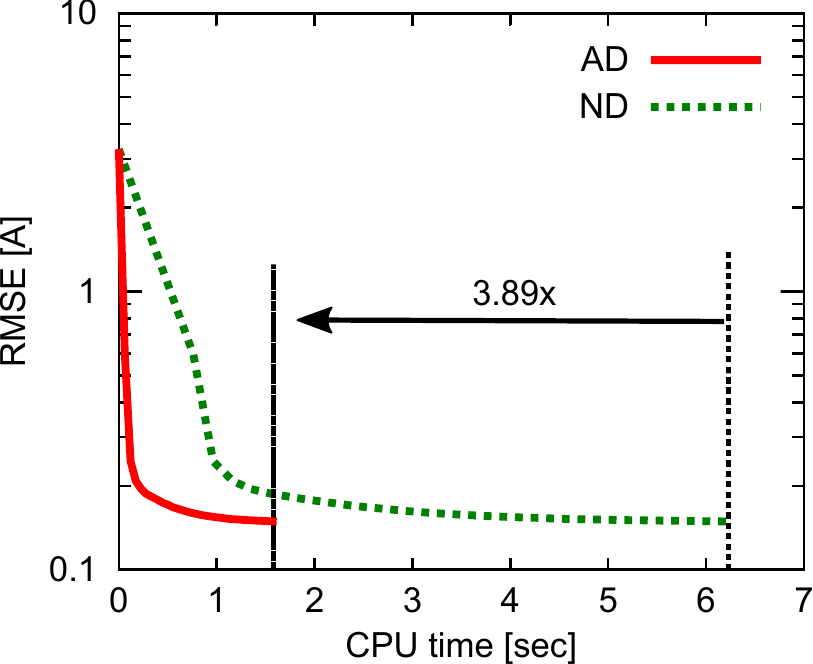}}
  \caption{RMSE reduction in the LM-based parameter extraction.}
  \label{fig:lm}
\end{figure}

Fig.~\ref{fig:lm} shows the RMSE reduction in the LM-based parameter
extraction. In this experiment, the model parameters of the
surface-potential-based model are extracted by using the same initial
parameters as those in Sec.~\ref{sec:c-1}.  Owing to the quadratic
convergence property of the LM algorithm~\cite{Computing2005-Fan}, the
number of iterations is reduced to 30 for both the AD- and ND-based
calculations, while the gradient-descent-based method requires more
than six hundred iterations as shown in Fig.~\ref{fig:lm_iter}.  As
expected, close to the ideal 3.89$\times$ acceleration has been
obtained, as shown in Fig.~\ref{fig:lm_time}.  In addition, the
extracted parameters by the AD- and ND-based calculations, which are
shown in Table~\ref{tab:lm_param}, are in perfect agreement, and quite
close to those in Table~\ref{tab:sp_error}.  From these results, the
proposed method can be successfully applied to other derivative-based
optimization algorithms, and the accelerations according to the number
of parameters can be achieved.

\begin{table}[!t]
  \centering
  \caption{Extracted values for the surface-potential-based model in the
    LM-based optimization}
  \label{tab:lm_param}
  \begin{tabular}{l||r|r|r} \hline
    Parameter&  &  & Relative  \\
    (Initial value) & AD & ND & error [\%]  \\ \hline
    {\bf SCALE} & & &  \\
    (5166360) & 516636 & 516636 & 0.000 \\
    {\bf TOX} & && \\
    ($5.0\times 10^{-8}$) & $4.626\times10^{-08}$ & $4.626\times10^{-08}$ & 0.000 \\
    {\bf NA} & &&\\
    ($1.31\times 10^{17}$) & $1.273\times10^{17}$ &$1.273\times10^{17}$& 0.000 \\
    {\bf LAMBDA} & &&\\
    ($8.69\times 10^{-3}$) & $5.312\times10^{-3}$ & $5.313\times10^{-3}$& 0.000 \\
    {\bf VFBC} & &&\\
    ($-4.90$) & $-1.597\times10^{-3}$ &$-1.597\times10^{-3}$ & 0.000  \\
    {\bf THETA} & &&\\
    ($5.91\times10^{-3}$) & $5.910\times10^{-8}$ &$5.910\times10^{-8}$ & 0.000 \\
    {\bf DELTA} & &&\\
    ($0.80$) & 0.614 & 0.614 & 0.000 \\
    {\bf RD} & &&\\ 
    ($2.90\times10^{-3}$) & $2.902\times10^{-3}$ &$2.902\times10^{-3}$ & 0.000  \\ \hline
  \end{tabular}
\end{table}

\subsection{Manufacturing variability analysis}
Recently, the application of artificial intelligence has been gaining
popularity in the area of power electronics~\cite{TPLE2021_Shuai}. As
an example, a neural network has been applied to model the device
current characteristics in SPICE simulators~\cite{TPLE2019_Chiozzi}
considering the parameter fitting as a black-box regression with a
neural network.  However, as opposed to the physics-based modeling,
neural networks carry out purely mathematical fitting of the measured
device characteristics to a nonlinear equation, ignoring physical
behavior of the device in interest. In order to fully understand
various device characteristics such as aging-induced threshold voltage
shift~\cite{TED2014_Rescher} and manufacturing process
variation~\cite{APEC2014_Wang,JESTPM2019_Borghese}, which have been
critical issues in the design of power converters using wide-bandgap
semiconductors, the usage of a set of physically meaningful model
parameters is of enormous importance.  Although the proposed method
incorporates a similar technique used in neural networks for
model-parameter extraction, our method preserves the physical meanings
of the model parameters.

\begin{figure}[!t]
  \centering
  \includegraphics[width=0.55\linewidth]{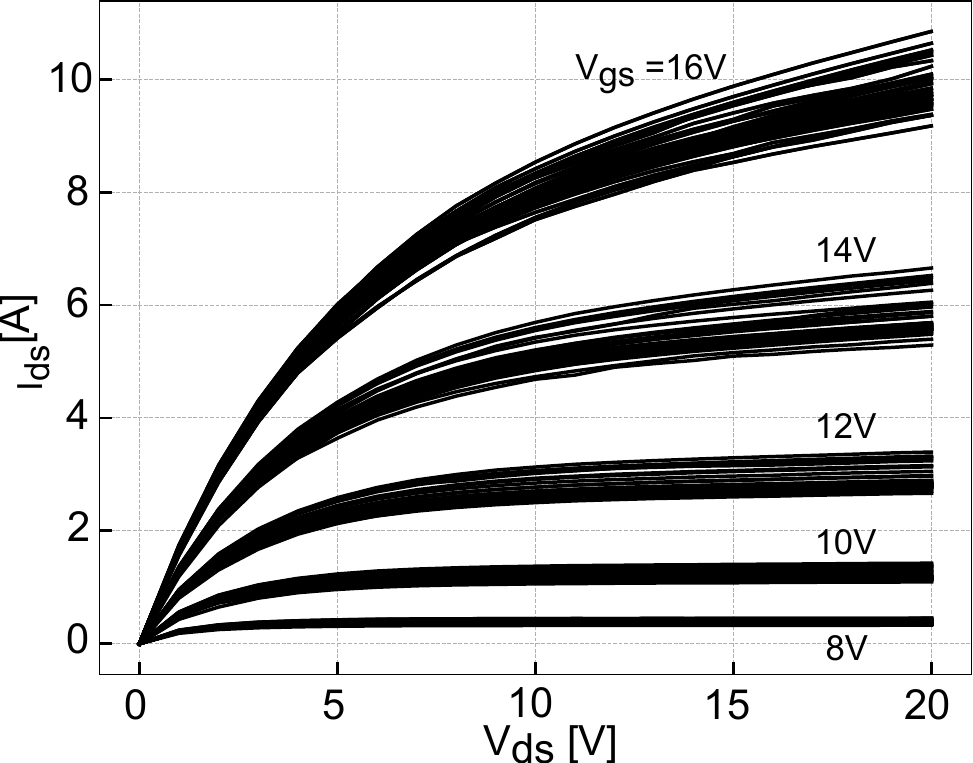}
  \caption{Measured I-V characteristics of 35 SiC MOSFETs.}
  \label{fig:var}
\end{figure}

\begin{figure}[!t]
  \centering
  \includegraphics[width=0.88\linewidth]{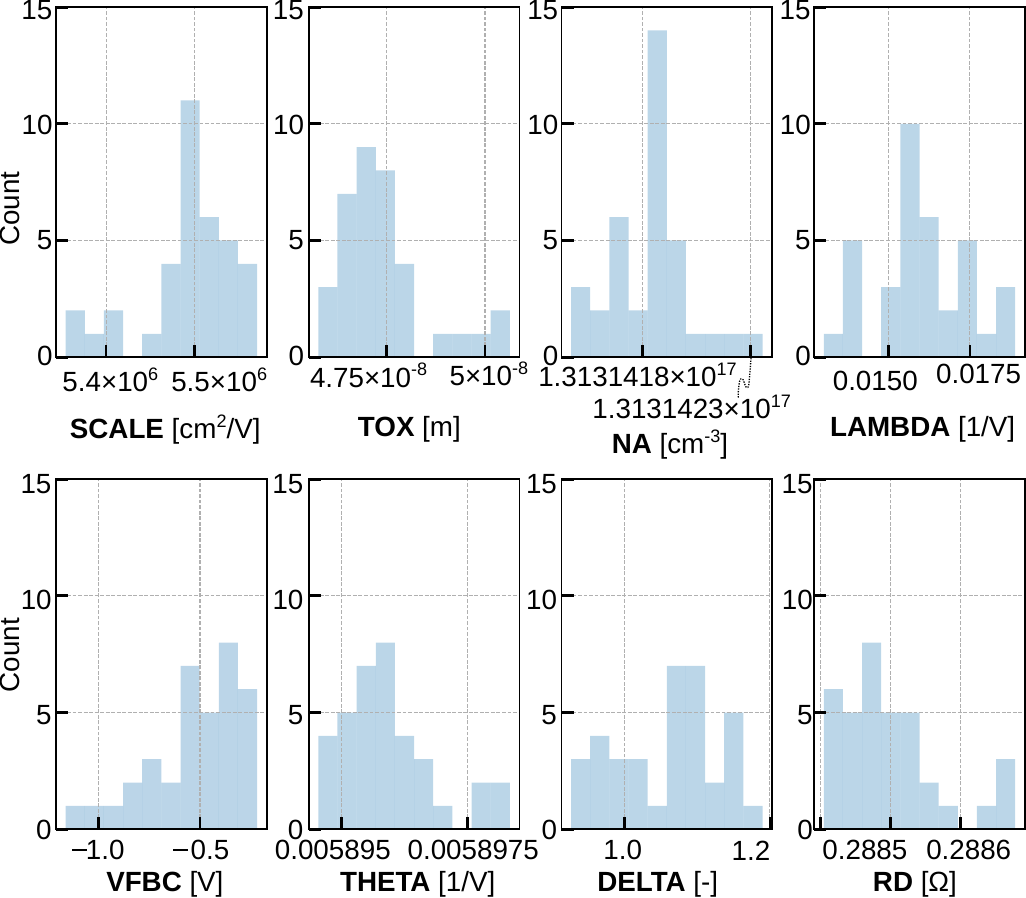}
  \caption{Histograms of the extracted model parameters of 35 SiC MOSFETs.}
  \label{fig:hist}
\end{figure}

In order to demonstrate the aforementioned advantages, we apply the
proposed method to the modeling of I-V characteristics of 35 SiC
MOSFETs, which are presented in Fig.~\ref{fig:var}.  For instance, in
the parallel operation of SiC MOSFETs in a power module, the
characteristics mismatch causes the current imbalances, resulting in
the significant degradation of the
reliability~\cite{APEC2014_Wang}. By applying the proposed method with
the surface-potential-based model, the distribution of its model
parameters can be obtained as shown in Fig.~\ref{fig:hist}.  We would
like to note that the extracted model parameters represent their
variation while the neural network based
method~\cite{TPLE2019_Chiozzi} only gives us 35 black box models
consisting of slightly different parameters. According
to~\cite{JESTPM2019_Borghese}, the current imbalance can be mitigated
by pairing MOSFETs with similar threshold voltage (${\bf VFBC}$)
and/or current gain factor (${\bf SCALE}$) in parallelly connected SiC
MOSFETs. Choosing pairs could be easy with the parameters obtained by
the proposed method.  Moreover, the distribution of the physically
meaningful parameters, such as {\bf TOX} and {\bf NA}, can also be
used by manufacturers to improve manufacturing yields.

\section{Conclusion}\label{sec:conclusion}

Herein, we proposed a novel parameter-extraction method that is applicable
to various power MOSFET models. The proposed method employs AD, which is
commonly used in backpropagation of artificial neural networks. The AD
technique involves the calculation of the partial derivatives according to
the chain rule by simply traversing the calculation graph, thereby reducing
the number of times the model equation must be evaluated compared to ND
calculation. Due to this improvement, the AD approach requires less
computation time than the ND approach for parameter extraction. Experimental
results using a commercial SiC power MOSFET demonstrated that the proposed
method could be used to successfully derive the model parameters for the
current equation of the MOSFET about 4.3 times faster than the conventional
gradient-descent method using two-point gradient approximations.

In future study, we intend to include an evaluation with open source
packages for AD as introduced in~\cite{autodiff} to analyze more
complex descriptions in arbitrary power MOSFET models by the
computational graph. In addition, the evaluation of the proposed
method is still limited in the static characteristics. We intend to
investigate a further evaluation on the dynamic characteristics based
on the extracted parameters.

\section*{Acknowledgment}
This work was partially supported by JST-OPERA Program Grant Number
JPMJOP1841, Japan.
The part of this work is also supported by JSPS KAKENHI Grant 20H04156 and 20K21793.

\section*{Appendix}
The proposed method can be applied even when the model contains loops, such
as ``while'' or ``for'' statements. The computational graph of the
loop is constructed with a feedback loop.  In the forward mode, the
computational graph is repeatedly traversed until the convergence
condition is satisfied.  Here, the variables of the graph are extended
to use a {\it stack} data structure so that the intermediate values
for each calculation are pushed into it. The backward mode is
conducted by popping the variables while traversing the graph. Here,
the while statement essentially contains a conditional branch.
Meanwhile, some models also contain ``if'' statement, so that
different calculations are carried out depending on the bias condition
or the parameter. In the proposed method, the conditional branch is
realized by adding a flag to the node again as a stack to judge which
path should be selected. The backward mode is carried out by
traversing the graph according to the flag value.

\begin{figure}[!t]
  \centering
  \includegraphics[width=0.65\linewidth]{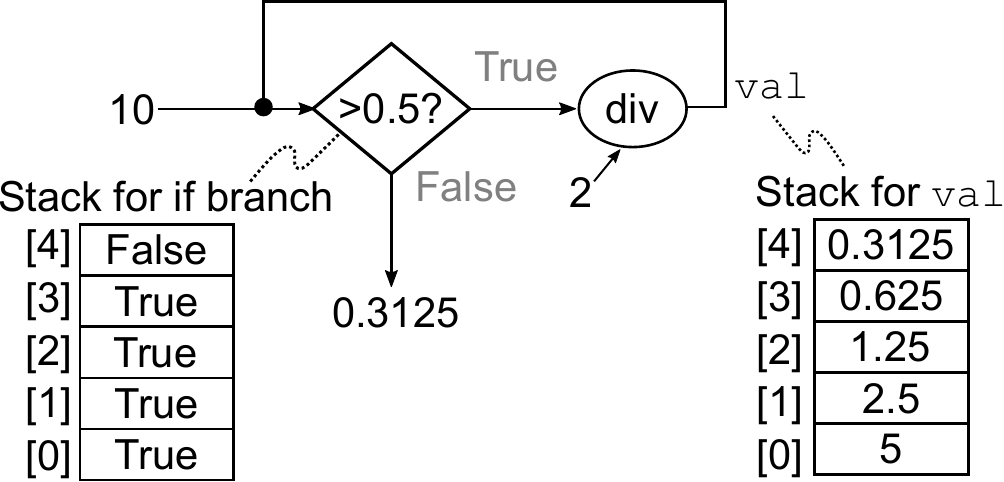}
  \caption{Example of a computational graph for a while loop.}
  \label{fig:loop}
\end{figure}

Fig.~\ref{fig:loop} shows a computational graph of a while statement
in the following code snippet as an example.
\begin{lstlisting}[basicstyle=\ttfamily]
  val = 10;
  while (val > 0.5){
    val = val / 2;
  }
\end{lstlisting}
In this code, variable {\tt val} is initialized to 10, then changes to
5, 2.5, 1.25, 0.625, and 0.3125.  Finally, the loop terminates when
its value becomes smaller than 0.5.  The computational graph of the
while statement is constructed with a feedback loop containing a
conditional branch and stacks as shown in Fig.~\ref{fig:loop}. In the
forward mode, the model equation is evaluated while pushing the value
of variable {\tt val} into the stack sequentially.  In addition, the
flags that will be used to which path should be evaluated are stored
in the stack.  Then, in the backward mode, the graph is traversed by
popping these values.

\ifCLASSOPTIONcaptionsoff
  \newpage
\fi



%

\bibliographystyle{IEEEtran}
\bibliography{alias,hvdev,cad}

%


\end{document}